\documentclass[
superscriptaddress,
showkeys,
showpacs,
amsmath,amssymb,
pre,
twocolumn,
]{revtex4-2}

\usepackage{lineno}
\usepackage[colorlinks,linkcolor=red,citecolor=blue,urlcolor=blue]{hyperref}
\usepackage{amsfonts}
\usepackage{amssymb}
\usepackage{float}
\usepackage{stmaryrd}
\usepackage{mathtools}

\usepackage[toc,page]{appendix}
\usepackage{graphicx}
\usepackage{epstopdf}
\usepackage{subcaption}
\usepackage{framed} 
\usepackage{nomencl} 
\usepackage{pgfplots}
\pgfplotsset{compat=1.14}
\usepackage{dcolumn}
\usepackage{bm}
\usepackage{soul}

\usepackage[draft]{todonotes}

\begin{document}
	\title{Entropic equilibrium for the lattice Boltzmann method:\\ Hydrodynamics and numerical properties}
	\author{S. A. Hosseini}
	\affiliation{Department of Mechanical and Process Engineering, ETH Zurich, 8092 Zurich, Switzerland}%
	\author{I. V. Karlin}\thanks{Corresponding author}
 \email{ikarlin@ethz.ch}
	\affiliation{Department of Mechanical and Process Engineering, ETH Zurich, 8092 Zurich, Switzerland}%
	\date{\today}
	\begin{abstract}
	    The entropic lattice Boltzmann framework proposed the construction of the discrete equilibrium by taking into consideration minimization of a discrete entropy functional. The effect of this form of the discrete equilibrium on properties of the resulting solver has been the topic of  discussions in the literature. Here we present a rigorous analysis of the hydrodynamics and numerics of the entropic. In doing so we demonstrate that the entropic equilibrium features unconditional linear stability, in contrast to the conventional polynomial equilibrium. We reveal the mechanisms through which unconditional linear stability is guaranteed, most notable of which the adaptive normal modes propagation velocity and the positive-definite nature of the dissipation rates of all eigen-modes. We further present a simple local correction to considerably reduce the deviations in the effective bulk viscosity.
	\end{abstract}
	\maketitle
\section{\label{sec:Introduction}Introduction}
The lattice Boltzmann method is a numerical method developed in the late 80's/early 90's, as an alternative to classical solvers for the Navier-Stokes equations, initially, in the incompressible flow limit~\cite{succi_lattice_2002}. This approach finds its roots in the kinetic theory of gases and is essentially a solver for the phase-space discretized Boltzmann equation with the linear Bhatnagar--Gross--Krook (BGK) approximation for the collision term~\cite{bhatnagar_model_1954}. The relatively uncomplicated algorithm, low cost of discrete operations in the time/space evolution equations, their locality stemming from the purely hyperbolic nature of the corresponding equations -as compared to elliptic-hyperbolic alternatives such as the Poisson--Navier--Stokes equations, and properties such as low numerical dissipation and strictly conservative nature have pushed the lattice Boltzmann method to the forefront of computational fluid dynamics.\\
Early on after its first appearance in the literature, numerical stability became a major point of discussion and research. In its original form, i.e. the single relaxation time BGK model with a quadratic-in-velocity discrete equilibrium, the lattice Boltzmann solver has been reported to be sensitive to the viscosity non-dimensionalized by the time-step and grid sizes, also referred to as the Fourier number, and the maximum non-dimensional velocity~\cite{kruger_lattice_2017}. Note that such limitations are not proper to the lattice Boltzmann method and come with any other numerical approximation to a system of hyperbolic/parabolic partial differential equations, as illustrated by the Lax equivalence theorem for finite difference methods~\cite{smith_numerical_1985,strikwerda_finite_2004} and the Lax-Richtmyer stability condition~\cite{lax_survey_1956}. Plethora of modifications to the original lattice BGK (LBGK) have been proposed since, the majority of which can be categorized as some form of multiple relaxation times putting in effect the burden of instabilities onto relaxation rates of individual independent moments of the distribution function~\cite{latt_lattice_2006,lallemand_theory_2003,geier_cascaded_2006,geier_cumulant_2015}. While such schemes have been successful, to different and limited extent, in extending the operation range of the LBGK, none of them has actually resulted in a scheme that is unconditionally stable in the linear regime. Based on that observation one might be tempted to ask whether focus should not be put on other components of the LBGK algorithm, for instance the discrete equilibrium. While the continuous equilibrium distribution function, i.e. Maxwell-Boltzmann distribution, is a minimizer of entropy and the Boltzmann-BGK equation complies with the H-theorem, it is not guaranteed that truncated expansions of the distribution function in a given projection space will abide by these conditions, i.e. non-commutativity of entropy minimization and discretization in phase-space. The entropic lattice Boltzmann method~\cite{karlin_maximum_1998} proposed a lattice Boltzmann realization guaranteeing stability based on a construction of the equilibrium attractor taking into account minimization of the discrete entropy. While the entropic lattice Boltzmann method has been successfully used for a wide variety of large Reynolds number simulations, see for instance~\cite{mazloomi_entropic_2015,chikatamarla_entropic_2006,prasianakis_lattice_2008}, the question of its stability as compared to other lattice Boltzmann realizations has been a topic of debates in the literature, see for instance~\cite{yong_nonexistence_2003,luo_numerics_2011,karlin_comment_2011,luo_reply_2012}, as best illustrated by the following quote from~\cite{luo_numerics_2011}: "... {it is unclear theoretically how the ELBE with a constant relaxation parameter} $\tau$ {can improve the numerical stability of the LBGK scheme}."\\
This sentence summarizes, in a nutshell, the question that will be considered in the present manuscript. Does the form of the discrete equilibrium distribution function, for a fixed relaxation scheme, affect the linear stability of the solver? Can the specific form of the discrete equilibrium derived by minimization of discrete entropy guarantee unconditional linear stability? And if so what are the mechanisms by which the entropic equilibrium guarantees linear stability? The present work will address these questions with a detailed analysis of different forms of the discrete equilibrium distribution function. Note that the present work only addresses the question of linear stability, as non-linear stability in the entropic lattice Boltzmann method is brought about by another ingredient which is dynamic restriction of the maximum path-length in the relaxation process.\\
The manuscript is organized as follows: after a brief review of basic concepts of the lattice Boltzmann method and different discrete equilibria in section \ref{sec1}, the continuum-level dynamics of the equivalent macroscopic balance equations will be analyzed. First dispersion properties will be dissected by looking into the Euler-level equations in section \ref{sec2}. Dissipation at the Navier-Stokes level will then be discussed in section \ref{sec3}. In section \ref{sec4} the analyses will be extended to the entire wave-number space through the von Neumann method and continuum limit results of sections \ref{sec3} and \ref{sec4} will be confirmed via numerical simulations. In addition, a simple and local correction is proposed to considerably reduce the deviations in the effective bulk viscosity for all equilibria. The manuscript will end with section \ref{sec5} summarizing all observation and discussing larger impact of the present study.
\section{Basic concepts\label{sec1}}
In the remainder of the present article, unless stated otherwise, we will consider single relaxation time lattice Boltzmann models with the following discrete time-evolution equation:
\begin{equation}
    f_i(\bm{r}+\bm{c}_i \delta t, t+\delta t) - f_i(\bm{r}, t)= 2\beta\left( f_i^{\rm eq}(\rho, \bm{u}) - f_i(\bm{r}, t)\right).
\end{equation}
Here $f_i$ are the discrete distribution functions, $\rm{r}$ the position in space, $t$ time, $\delta t$ the time-step size, $\rho$ the fluid density, $\bm{u}$ the velocity and $\beta$ the relaxation frequency defined as~\cite{ansumali_stabilization_2000,ansumali_entropy_2002}:
\begin{equation}\label{eq:relaxation_frequency}
    \beta = \frac{\delta t}{2\nu/\varsigma^2 + \delta t},
\end{equation}
where $\nu$ is the fluid kinematic viscosity and $\varsigma$ the lattice sound speed:
\begin{equation}
    \varsigma = \frac{\delta r}{\sqrt{3}\delta t},
\end{equation}
and $\delta r$ the grid size. $f_i^{\rm eq}$ are the equilibrium distribution functions which will be the focus of this work. Here all analyses and discussions will consider first-neighbor lattices, in 1-D,
\begin{equation}
    c_i\in\frac{\delta r}{\delta t}\{-1,0,1\},
\end{equation}
and corresponding tensorial products in 2- and 3-D. The weights associated with this lattice are, in 1-D,
\begin{equation}
    w_i\in\left\{\frac{1}{6},\frac{2}{3},\frac{1}{6}\right\}.
\end{equation}
Weights for 2- and 3-D lattices can be obtained as products of 1-D weights.\\
Details about the equilibrium distribution functions, starting with the entropic construction are given in the next subsections.
\subsection{Discrete entropic equilibrium construction}
In the context of the entropic lattice Boltzmann method the discrete equilibrium state is found as the minimizer of a Lyapunov functional $H$~\cite{karlin_perfect_1999,karlin_maximum_1998}:
\begin{equation}
    H = \sum_{i=1}^{Q} h_i(f_i),
\end{equation}
where $h_i$ are convex functions, under mass and momentum conservation constraints:
\begin{align}
	\sum_{i=1}^Q f_i^{\rm eq} = \rho,\\
    \sum_{i=1}^Q \bm{c}_i f_i^{\rm eq} = \rho \bm{u},
\end{align}
which is different from the polynomial construction which also imposes a constraint on the second-order moment of the equilibrium distribution function. Introducing the corresponding Lagrange multipliers, $\lambda_0, \lambda_\alpha$:
\begin{equation}
    \delta \sum_{i=1}^{Q} \left( h_i(f_i) - \lambda_0 f_i - \lambda_\alpha c_{i\alpha} f_i \right) = 0,
\end{equation}
which yields
\begin{equation}
    h_i'(f_i^{\rm eq}) = \lambda_0 + \lambda_\alpha c_{i\alpha} .
\end{equation}
Defining the inverse of $h_i'(f_i^{\rm eq})$ as $\mu_i = \left[{h_i'(f_i^{\rm eq})}\right]^{-1}$, which must exist due to the convexity of $h_i$ the formal solution of the minimization problem reads:
\begin{equation}
    f_i^{\rm eq} = \mu_i \left( \lambda_0 + \lambda_\alpha c_{i\alpha} \right).
\end{equation}
Taking the $H$-function to be~\cite{ansumali_minimal_2003}:
\begin{equation}
    H = \sum_{i=1}^{Q} f_i \ln(f_i/w_i),
\end{equation}
and considering stencils with $3^D$ discrete velocities the corresponding equilibria can be expressed as:
    \begin{equation}
    f_i^{\rm eq} = w_i \exp\left(\lambda_0\right) \prod_{\alpha=1}^{D}\exp\left(c_{i\alpha}\lambda_\alpha\right).
    \end{equation}
Introducing the following changes of variables, $X = \exp\left(-\lambda_0\right)$ and $Z_\alpha = \exp\left(\lambda_\alpha\right)$ the equilibrium can be re-written as:
    \begin{equation}
    f_i^{\rm eq} = w_i X^{-1} \prod_{\alpha=1}^{D} Z_\alpha^{c_{i\alpha}}.
    \end{equation}
Here $X$ and $Z_\alpha$, i.e. the corresponding Lagrange multipliers, can be obtained by writing the constraints on the moments:
\begin{subequations}
	\begin{align}
		\rho X &= \sum_{i=1}^{Q} w_i \prod_{\alpha=1}^{D} Z_\alpha^{c_{i\alpha}}, \\
		\rho u_\beta X &= \sum_{i=1}^{Q} w_i c_{i\beta} \prod_{\alpha=1}^{D} Z_\alpha^{c_{i\alpha}}.
	\end{align}
\end{subequations}
Solving this system of equations results in:
\begin{equation}
    Z_\alpha = \frac{2u_\alpha + \sqrt{{\left(\frac{u_\alpha}{\varsigma}\right)}^2 + 1}}{1-u_\alpha},
\end{equation}
\begin{equation}
    X^{-1} = \rho \prod_{\alpha=1}^{D} \left(2-\sqrt{{\left(\frac{u_\alpha}{\varsigma}\right)}^2 + 1}\right),
\end{equation}
leading to the following final expression for the discrete entropic equilibrium~\cite{ansumali_minimal_2003}:
    \begin{equation}\label{eq:Entropic_isothermal_EDF}
    f_i^{\rm eq} =  w_i \rho  \prod_{\alpha=1}^{D} \left(2-\sqrt{{\left(\frac{u_\alpha}{\varsigma}\right)}^2 + 1}\right) {\left(\frac{2u_\alpha + \sqrt{{\left(\frac{u_\alpha}{\varsigma}\right)}^2 + 1}}{1-u_\alpha}\right)}^{c_{i\alpha}}.
    \end{equation}
\subsection{Polynomial discrete equilibria}
Another class of discrete equilibria, widely used in discrete velocity approaches, are the polynomial equilibria. The most well-known realization of polynomial equilibria for isothermal flows is the second-order polynomial equilibrium defined as~\cite{kruger_lattice_2017}:
\begin{equation}\label{eq:Poly2_isothermal_EDF}
    f_i^{\rm eq} = w_i \rho\left( 1 + \frac{\bm{c}_i\cdot\bm{u}}{\varsigma^2} + \frac{{\left(\bm{c}_i\cdot\bm{u}\right)}^2}{2\varsigma^4} - \frac{\bm{u}^2}{2\varsigma^2}\right).
\end{equation}
While initially derived via a second-order Taylor expansion of the Maxwell-Boltzmann distribution around Ma=0, it can also be obtained as a second-order Hermite expansion of the Maxwell-Boltzmann distribution function.\\
This form of the equilibrium is known to lead to Galilean-variant errors in the off-diagonal components of the viscous stress tensor, which scale with order three in Mach number when the local temperature coincides with the lattice reference temperature and order one when the local temperature is different from the lattice reference. Attempts at improving upon the second-order polynomial equilibrium, especially for compressible flows, have led to the product form equilibrium:
\begin{equation}\label{eq:Product_isothermal_EDF}
    f_i^{\rm eq} = \rho \prod_{\alpha=1}^{D} {\left(1-\varsigma^2-u_{\alpha}^2\right)}^{1-\lvert c_{i\alpha}\lvert} {\left( \frac{c_{i\alpha} u_\alpha + \varsigma^2 + u_\alpha^2}{2}\right)}^{\lvert c_{i\alpha}\lvert}.
\end{equation}
This form of the equilibrium was obtained in \cite{prasianakis_lattice_2007} by \emph{guiding} the minimization problem of the previous section by explicitly adding a constraint on diagonal components of the second-order equilibrium moment tensor. This form of the equilibrium can also be obtained, in 2-D, via a fourth order Hermite expansion~\cite{shan_discretization_1998}, or the moment matching method~\cite{hosseini_compressibility_2020}.\\

\section{Euler-level dynamics: dispersion\label{sec2}}
In this section we will discuss Euler level dynamics of the hydrodynamic equations corresponding to each discrete equilibrium distribution function.
\subsection{Conservation equations}
\begin{figure*}
    \centering	    \includegraphics[width=11cm,keepaspectratio]{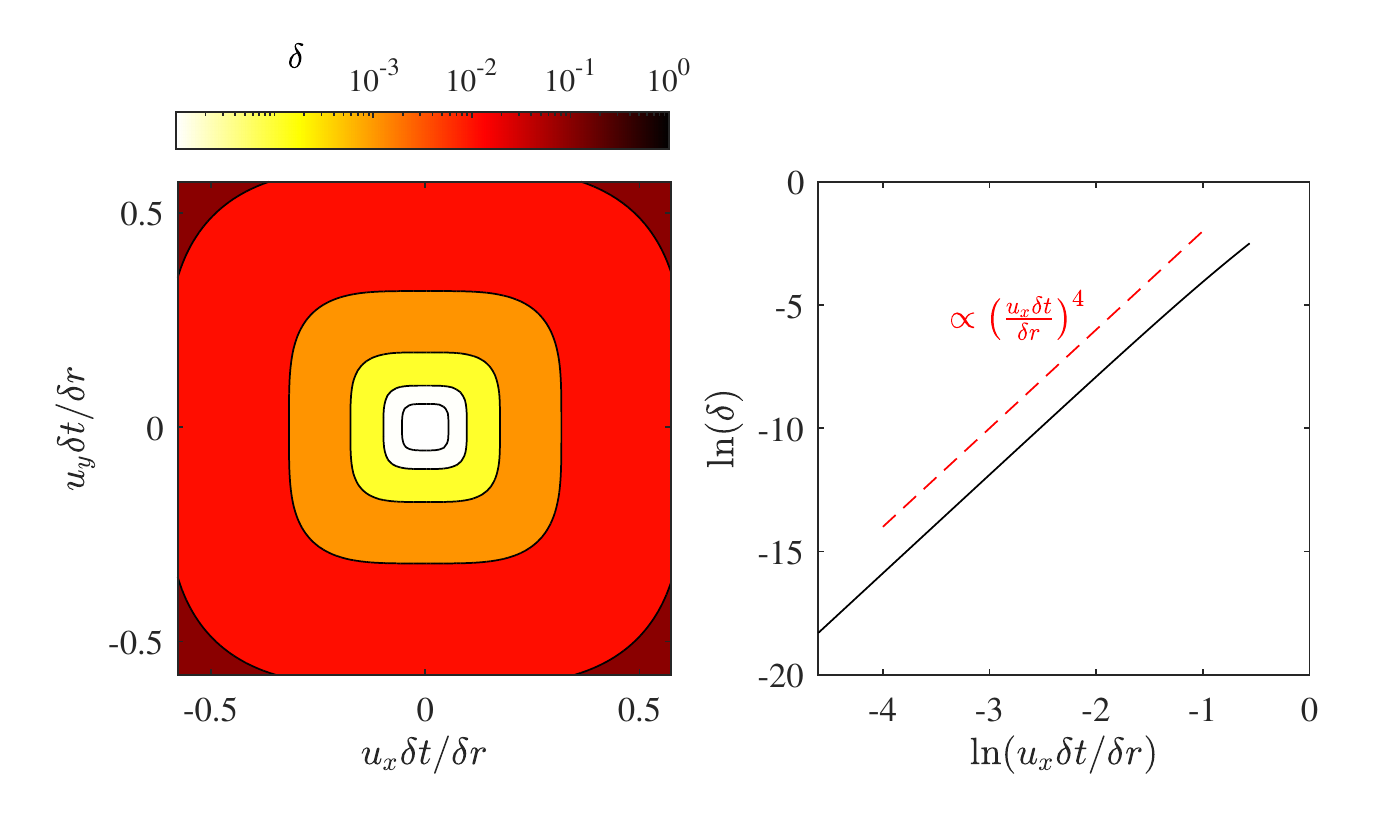}
    \caption{Illustration of deviations in trace of pressure tensor for the entropic equilibrium. Left: Normalized deviation in 2-D. Right: Normalized error in 1-D.}
    \label{Fig:entropic_eq_Pxx_Pyy}
\end{figure*}

Applying the Chapmann-Enkog analysis to the entropic lattice Boltzmann equations, at order $\varepsilon$ one finds the following conservation laws:
\begin{subequations}
    \begin{align}
        \partial_t^{(1)}\rho + \bm{\nabla}\cdot\rho\bm{u} &= 0, \label{eq:continuity}\\
        \partial_t^{(1)}\rho\bm{u} + \bm{\nabla}\cdot\rho\bm{u}\otimes\bm{u} + \bm{\nabla}\cdot\rho\varsigma^2\bm{I} + \bm{\nabla}\cdot \bm{P}^{*} &= 0.\label{eq:euler}
    \end{align}
\end{subequations}
Here $\bm{P}^{*}$ is a diagonal matrix defined as:
\begin{equation}
    P^{*}_{\alpha\alpha} = \widetilde{\Pi}_{\alpha\alpha}^{\rm eq} - \rho \varsigma^2 = \sum_{i=1}^{Q} {(\bm{c}_i-\bm{u})}^2 f^{\rm eq}_i - \rho \varsigma^2,
\end{equation}
Using Eq.~\eqref{eq:Entropic_isothermal_EDF}, $P^{*}_{\alpha\alpha}$ is readily computed:
\begin{equation}
    P^{*}_{\alpha\alpha} = -\rho \varsigma^2 \left[{(u_\alpha/\varsigma)}^2 - 2\sqrt{{(u_\alpha/\varsigma)}^2 + 1} + 2\right],
\end{equation}
while for Eqs.~\eqref{eq:Poly2_isothermal_EDF} and \eqref{eq:Product_isothermal_EDF} $P^{*}_{\alpha\alpha}=0$.
As defined here, $P^{*}_{\alpha\alpha}$ is the deviation of the pressure from the Maxwell--Boltzmann pressure, which is both flow-dependent and anisotropic for the entropic equilibrium. Note that in the limit of $u_\alpha\rightarrow\pm\delta r/\delta t$ the deviation, $P^*_{\alpha\alpha}$ goes to $-\rho\varsigma^2$ resulting in a vanishing total thermodynamic pressure.
The trace of the deviation is illustrated in Fig.~\ref{Fig:entropic_eq_Pxx_Pyy} via the following normalized variable:
\begin{equation}
    \delta = \frac{P^{*}_{xx} + P^{*}_{yy}}{2\rho \varsigma^2}.
\end{equation}
Based on Fig.~\ref{Fig:entropic_eq_Pxx_Pyy}, as is well known, the deviation of pressure implied by the entropic equilibrium are immaterial for nearly-incompressible flow simulations. However, as we shall see in the analysis below, precisely these deviations have crucial implications on the well-posedness and stability of LBGK with the entropic equilibrium.
\subsection{Characteristics analysis: Eigen-modes}
In the case of polynomial equilibria of order equal to or larger than two the second-order central moment, representing pressure, is only a function of local density and temperature:
\begin{equation}
    \widetilde{\Pi}_{\alpha\alpha} = \rho \varsigma^2,
\end{equation}
and the corresponding sound speed, $c_s$, can be readily computed via:
\begin{equation}\label{eq:csrelative}
    c_s = u_\alpha \pm \sqrt{\frac{\partial \widetilde{\Pi}_{\alpha\alpha}}{\partial \rho}\bigg|_{T={\rm cst}}} = u_\alpha \pm \varsigma.
\end{equation}
In the case of the entropic equilibrium the equilibrium pressure is also function of local velocity making computation of the sound speed less evident. To compute an analytical expression for the sound speed with the entropic equilibrium, we  use the method of characteristics, briefly introduced in the next section.
\subsubsection{Characteristics analysis formalism}
Consider a system of conservation equations consisting of first-order partial differential equations of the following form~\cite{thompson_time_1987}:
\begin{equation}
    \partial_t \Phi + \partial_x \mathcal{F} + \mathcal{C} = 0,
\end{equation}
where $\Phi$ is the vector of conserved variables, $\mathcal{F}$ the vector of fluxes and $\mathcal{C}$ represents non-homogeneous terms in the equations. This system, which is represented above with the conservative form, can also be represented using the primitive form as:
\begin{equation}\label{eq:primitive_form}
    \partial_t \phi + \mathcal{A}\partial_x \phi + \mathcal{C}' = 0,
\end{equation}
where $\phi$ is the vector of primitive variable and $\mathcal{A}$ is a matrix of size $n\times n$ -for a system with $n$ independent variables. Note that unlike conserved variables, the choice of primitive variable is not unique. The two descriptions are related to eachother via:
\begin{subequations}
    \begin{align}
        \partial_t \Phi &= \mathcal{P} \partial_t \phi,\\
        \partial_x \mathcal{F} &= \mathcal{Q} \partial_x \phi,
    \end{align}
\end{subequations}
with:
\begin{subequations}
    \begin{align}
    \mathcal{P}_{mn} &= \frac{\partial \Phi_m}{\partial \phi_n}\\
    \mathcal{Q}_{mn} &= \frac{\partial \mathcal{F}_m}{\partial \phi_n},
    \end{align}
\end{subequations}    
and:
\begin{subequations}
    \begin{align}
        \mathcal{A} &= \mathcal{P}^{-1} \mathcal{Q},\\
        \mathcal{C} &= \mathcal{P}^{-1} \mathcal{C}'.
    \end{align}
\end{subequations}
Let us now introduce the left, $\bm{l}$, and right, $\bm{r}$, eigen-vectors of $\mathcal{A}$ such that:
\begin{subequations}
    \begin{align}
        l_m \mathcal{A} &= \lambda_m l_m,\\
        \mathcal{A} r_m &= \lambda_m r_m.
    \end{align}
\end{subequations}
where $\lambda_i$ are the eigen-values of $\mathcal{A}$. Then a diagonal matrix $\Lambda$ can be obtained via:
\begin{equation}
    \Lambda = \mathcal{S}\mathcal{A}\mathcal{S}^{-1},
\end{equation}
where the rows of $\mathcal{S}$ are the left eigen-vectors and the columns of $\mathcal{S}^{-1}$ are the right eigen-vectors. Multiplying Eq.~\eqref{eq:primitive_form} by $\mathcal{S}$ one obtains the characteristics form of the system:
\begin{equation}
    l_m \partial_t \phi + \lambda_m l_m\partial_x \phi + l_m \mathcal{C}' = 0,
\end{equation}
which upon introduction of a new variable $d\mathcal{V}_m = l_m d\phi + l_m \mathcal{C}'dt$ reduces to a system of wave equations with velocities $\lambda_m$:
\begin{equation}
    \partial_t \mathcal{V}_m + \lambda_m\partial_x \mathcal{V}_m = 0.
\end{equation}
This analysis can be readily extended to multiple physical dimensions, considering propagation along the $\alpha$-axis, by re-writing Eq.~\eqref{eq:primitive_form} as:
\begin{equation}\label{eq:primitive_form_mult_D}
    \partial_t \phi + \mathcal{A}_\alpha \partial_\alpha \phi + \mathcal{C}' = 0,
\end{equation}
where $\alpha\in\{x,y,z\}$.
\subsubsection{Characteristics: 1-D system}
Let us first apply this analysis to the system of hyperbolic conservation equations recovered by the D1Q3 lattice, shown in Eqs.~\eqref{eq:continuity} and \eqref{eq:euler}. The primitive form of this system will results in:
\begin{equation}
    \phi = \begin{bmatrix} \rho \\ u_x\end{bmatrix}
\end{equation}
and:
\begin{equation}
    \mathcal{A} = \begin{bmatrix} u_x & \rho \\ \varsigma^2 + \frac{\partial_{\rho} P^{*}_{xx}}{\rho} & u_x + \frac{\partial_{u_x} P^{*}_{xx}}{\rho}\end{bmatrix}.
\end{equation}
Applying the characteristics analysis to this system the following eigen-values are recovered:
\begin{subequations}
    \begin{align}
        c_s^{+} &= u_x + \frac{\partial_{u_x} P^{*}_{xx}}{2\rho}
        + \sqrt{{\left(\frac{\partial_{u_x}  P^{*}_{xx}}{2\rho}\right)}^2+{\varsigma}^2+ \partial_{\rho}  P^{*}_{xx}},\\
        c_s^{-} &= u_x + \frac{\partial_{u_x} P^{*}_{xx}}{2\rho}
        - \sqrt{{\left(\frac{\partial_{u_x}  P^{*}_{xx}}{2\rho}\right)}^2+{\varsigma}^2+ \partial_{\rho}  P^{*}_{xx}}.
    \end{align}
\end{subequations}
Using the left eigen-vectors, readily obtained as right eigen-vectors of $\mathcal{A}^{\dagger}$, the following wave system is recovered:
\begin{subequations}
    \begin{align}
    \left[\frac{c_s^{+}-u_x}{\rho}\partial_t \rho + \partial_t u_x \right] + c_s^{+}\left[\frac{c_s^{+}-u_x}{\rho}\partial_x \rho + \partial_x u_x \right] &= 0, \label{eq:cs_plus_1d}\\
    \left[\frac{c_s^{-}-u_x}{\rho}\partial_t \rho + \partial_t u_x \right] + c_s^{-}\left[\frac{c_s^{-}-u_x}{\rho}\partial_x \rho + \partial_x u_x \right] &= 0.\label{eq:cs_minus_1d}
    \end{align}
\end{subequations}
Note that for polynomial equilibria of Eqs.~\eqref{eq:Poly2_isothermal_EDF} and \eqref{eq:Product_isothermal_EDF}, i.e. $P^{*}_{xx}=0$, one recovers, in accord with \eqref{eq:csrelative}:
\begin{subequations}
    \begin{align}
        c_s^{+} &= u_x + \varsigma, \label{eq:cs_plus_poly}\\
        c_s^{-} &= u_x - \varsigma, \label{eq:cs_minus_poly}
    \end{align}
\end{subequations}
and
\begin{subequations}
    \begin{align}
    \left[\frac{\varsigma}{\rho}\partial_t \rho + \partial_t u_x \right] + \left(u_x+\varsigma\right)\left[\frac{\varsigma}{\rho}\partial_x \rho + \partial_x u_x \right] &= 0, \\
    \left[\frac{\varsigma}{\rho}\partial_t \rho - \partial_t u_x \right] + \left(u_x-\varsigma\right)\left[\frac{\varsigma}{\rho}\partial_x \rho - \partial_x u_x \right] &= 0.
    \end{align}
\end{subequations}
We will discuss the specific case of the entropic equilibrium in the next paragraphs.
\subsubsection{Characteristics: multi-dimensional systems}
Now for systems of dimension larger than one, here to illustrate a 2-D system:
\begin{equation}
    \phi = \begin{bmatrix} \rho \\ u_x\\ u_y \end{bmatrix}
\end{equation}
and:
\begin{equation}
    \mathcal{A}_x = \begin{bmatrix} u_x & \rho & 0 \\ \varsigma^2 + \frac{\partial_{\rho} P^{*}_{xx}}{\rho} & u_x + \frac{\partial_{u_x} P^{*}_{xx}}{\rho} & \frac{\partial_{u_y} P^{*}_{xx}}{\rho} \\
    0 & 0 & u_x
    \end{bmatrix}.
\end{equation}
and:
\begin{equation}
    \mathcal{A}_y = \begin{bmatrix} u_y & 0 & \rho \\ 0 & u_y & 0 \\ \varsigma^2 + \frac{\partial_{\rho} P^{*}_{y}}{\rho} & \frac{\partial_{u_x} P^{*}_{yy}}{\rho} & u_y + \frac{\partial_{u_y} P^{*}_{yy}}{\rho}\end{bmatrix}.
\end{equation}
and considering propagation along the $x-$axis one recovers the two acoustic modes of Eqs.~\eqref{eq:cs_plus_1d} and \eqref{eq:cs_minus_1d} and an additional \emph{shear} mode propagating at speed $u_x$. Applying these eigen-values and the left side eigen-vectors the following wave system is recovered:
\begin{subequations}
    \begin{align}
    \partial_t u_y + u_x \partial_x u_y &= 0, \label{eq:shear_2dx}\\
    \left[\frac{c_s^{+}-u_x}{\rho}\partial_t \rho + \partial_t u_x \right] + c_s^{+}\left[\frac{c_s^{+}-u_x}{\rho}\partial_x \rho + \partial_x u_x \right] &= 0, \label{eq:cs_plus_2dx}\\
    \left[\frac{c_s^{-}-u_x}{\rho}\partial_t \rho + \partial_t u_x \right] + c_s^{-}\left[\frac{c_s^{-}-u_x}{\rho}\partial_x \rho + \partial_x u_x \right] &= 0.\label{eq:cs_minus_2dx}
    \end{align}
\end{subequations}
Along the $y$-axis the wave system changes into:
\begin{subequations}
    \begin{align}
    \partial_t u_x + u_y \partial_y u_x &= 0, \label{eq:shear_2dy}\\
    \left[\frac{c_s^{+}-u_y}{\rho}\partial_t \rho + \partial_t u_y \right] + c_s^{+}\left[\frac{c_s^{+}-u_y}{\rho}\partial_y \rho + \partial_y u_y \right] &= 0, \label{eq:cs_plus_2dy}\\
    \left[\frac{c_s^{-}-u_y}{\rho}\partial_t \rho + \partial_t u_y \right] + c_s^{-}\left[\frac{c_s^{-}-u_y}{\rho}\partial_y \rho + \partial_y u_y \right] &= 0.\label{eq:cs_minus_2dy}
    \end{align}
\end{subequations}
where the acoustic propagation modes in $y-$direction are now:
\begin{subequations}
    \begin{align}
        c_s^{+} &= u_y + \frac{\partial_{u_y} P^{*}_{yy}}{2\rho}
        + \sqrt{{\left(\frac{\partial_{u_y}  P^{*}_{yy}}{2\rho}\right)}^2+{\varsigma}^2+ \partial_{\rho}  P^{*}_{yy}},\label{eq:cs_plus_gen}\\
        c_s^{-} &= u_y + \frac{\partial_{u_y} P^{*}_{yy}}{2\rho}
        - \sqrt{{\left(\frac{\partial_{u_y}  P^{*}_{yy}}{2\rho}\right)}^2+{\varsigma}^2+ \partial_{\rho}  P^{*}_{yy}}\label{eq:cs_minus_gen}.
    \end{align}
\end{subequations}
Analysis of the propagation speed of different eigen-modes shows that acoustic modes will propagate isotropically only in the limit of:
\begin{equation}
    \partial_\rho P_{xx}^{*} = \partial_\rho P_{yy}^{*},
\end{equation}
and
\begin{equation}
    \partial_{u_x} P_{xx}^{*} = \partial_{u_y} P_{yy}^{*}.
\end{equation}
A special solution of these conditions is $P_{xx}^{*}=P_{yy}^{*}=0$ which corresponds to the polynomial equilibria explicitly enforcing the second-order moment of the equilibrium distribution function, i.e. Eqs.~\eqref{eq:Poly2_isothermal_EDF} and \eqref{eq:Product_isothermal_EDF}. However these equilibria come with limitations that the entropic equilibrium does not have. This point will be discussed in the next paragraph.
\subsection{Stabilization of entropic equilibrium via sound speed}
\begin{figure*}
	\centering
	\includegraphics[width=12cm,keepaspectratio]{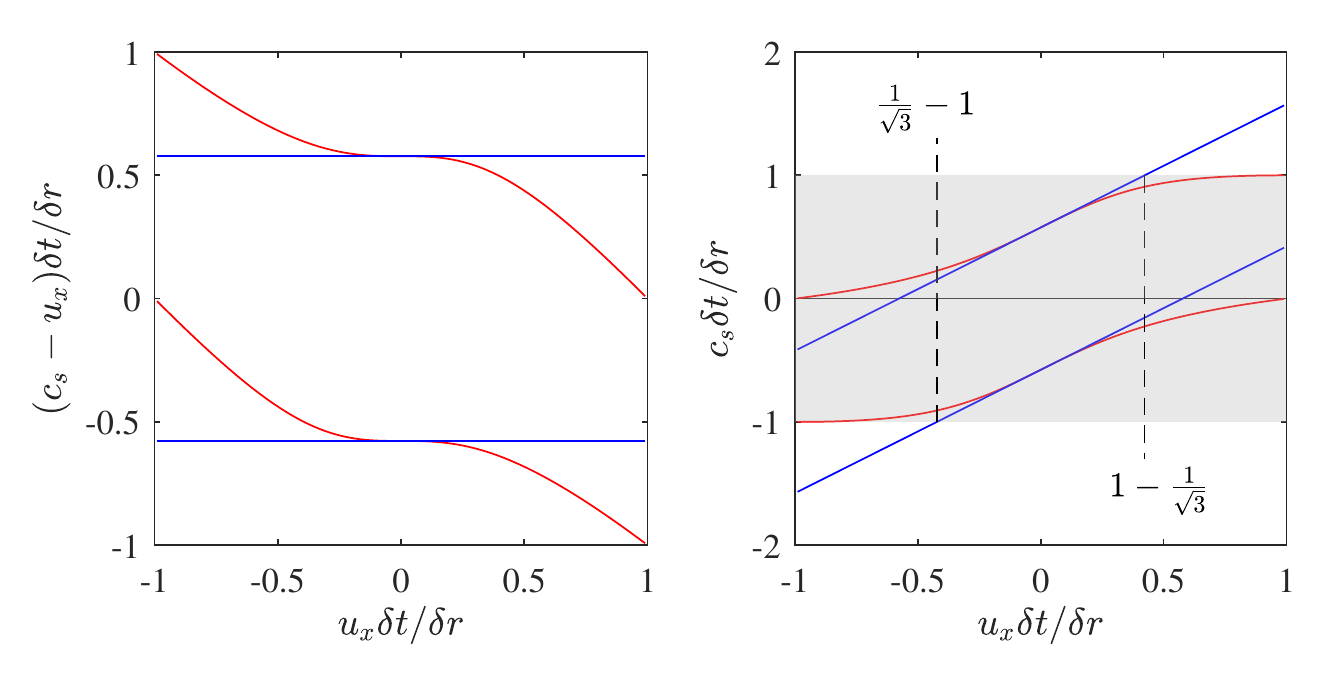}
	\caption{(Left) Sound speed for entropic, Eqs.~\eqref{eq:cs_plus_ent} and \eqref{eq:cs_minus_ent}, and polynomial equilibria, Eqs.~\eqref{eq:cs_plus_poly} and \eqref{eq:cs_minus_poly}, as a function of velocity $u_x$. (Right) Comparison of the speed of fastest propagating eigen-modes: (blue lines) polynomial and (red lines) entropic equilibria.}
	\label{Fig:sound_speed_entropic_vs_poly}
\end{figure*}

\begin{figure}[h!]
	\centering
	\includegraphics[width=6cm,keepaspectratio]{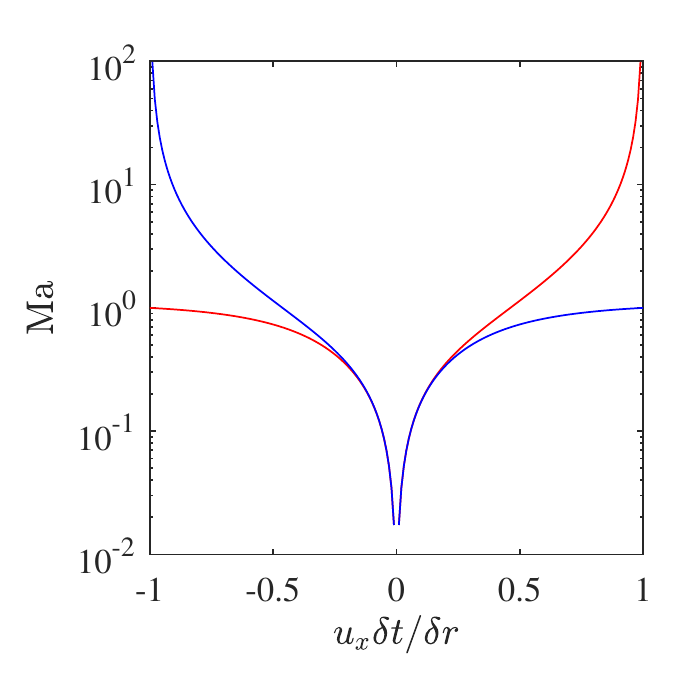}
	\caption{Mach number as a function of speed, $u_x$ as defined in Eqs.~\eqref{eq:entropic:map} and \eqref{eq:entropic:man} respectively in red and blue.}
	\label{Fig:entropic_mach_number}
\end{figure}

\begin{figure}[h!]
	\centering
	\includegraphics[width=6cm,keepaspectratio]{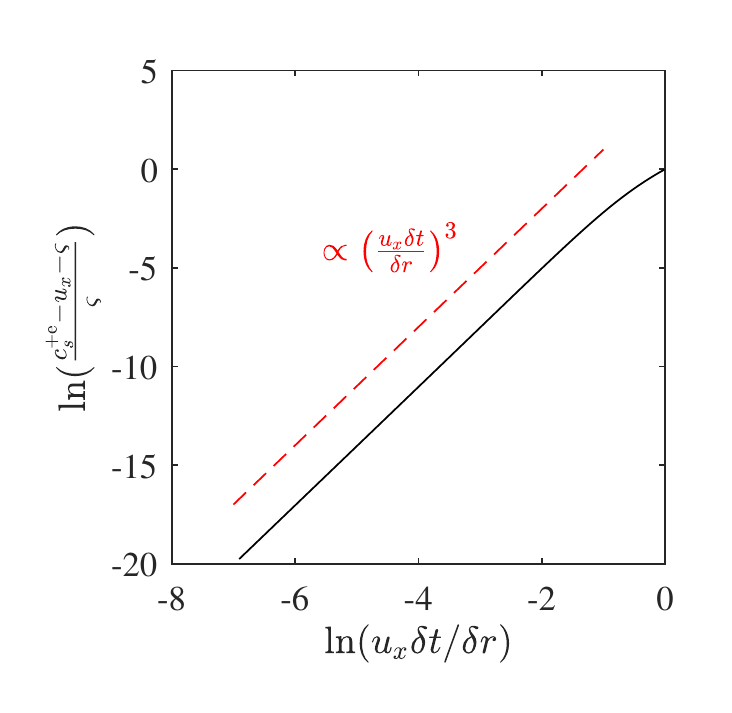}
	\caption{Normalized deviation of entropic equilibrium sound speed as a function of non-dimensional velocity.}
	\label{Fig:error_in_cs_entropic}
\end{figure}

Plugging the entropic equilibrium into the eigen-values obtained for the general formulation two non-symmetrical sound propagation speeds are recovered:
\begin{subequations}
\begin{align}
    c_s^{\rm e+} &= \frac{u_x + \varsigma\,\sqrt{2\,\sqrt{{\left(\frac{u_x}{\varsigma}\right)}^2+1}-1}}{\sqrt{{\left(\frac{u_x}{\varsigma}\right)}^2+1}}, \label{eq:cs_plus_ent}\\
    c_s^{\rm e-} &= \frac{u_x - \varsigma\,\sqrt{2\,\sqrt{{\left(\frac{u_x}{\varsigma}\right)}^2+1}-1}}{\sqrt{{\left(\frac{u_x}{\varsigma}\right)}^2+1}}.\label{eq:cs_minus_ent}
\end{align}
\end{subequations}
The propagation speed of these two eigen-modes are shown in Fig.~\ref{Fig:sound_speed_entropic_vs_poly} as a function of speed.
The behavior of entropic sound speed shows an interesting property of the entropic model pointing already to a (potentially) unconditional linear stability. Having re-written the equivalent Euler system in terms of coupled wave equations, note that for the solver to access information required to form the solution in time the numerical domain of dependence of any point in space and time must include the analytical domain of dependence, i.e. the initial conditions have an effect on the exact value of the solution at that point~\cite{courant_uber_1928}. Simply put, the fastest eigen-modes in the system \emph{cannot} propagate faster than the lattice links. Only considering three \emph{physical} eigen-modes, i.e. $u_x$, $c_s^{+}$ and $c_s^{-}$, one arrives at the following condition on linear stability:
\begin{equation}\label{eq:CFL_strict_condition}
    \max(\lvert u_x \lvert, \lvert c_s^{+} \lvert, \lvert c_s^{-} \lvert)\leq\frac{\delta r}{\delta t}.
\end{equation}
For the polynomial equilibria, given that sound speed is constant one recovers the following maximum tolerable velocity:
\begin{equation}\label{eq:CFL_lmit_absolute_poly}
    \lvert u_x^{\rm max} \lvert = \frac{\delta r}{\delta t}-\varsigma=0.4226\frac{\delta r}{\delta t},
\end{equation}
which as will be shown in the next section through stability analyses, is indeed the maximum reachable velocity. For the entropic equilibrium on the other hand, it is observed that the sound speed self-adjusts as a function of local velocity to guarantee Eq.~\eqref{eq:CFL_strict_condition} is always satisfied, leading to:
\begin{equation}
    \lvert u_x^{\rm max}\lvert = \frac{\delta r}{\delta t}.
\end{equation}
At the higher/lower end of the velocity spectrum, 
\begin{subequations}
    \begin{align}
        \lim_{u_x \rightarrow -\delta r/\delta t} c_s^{e+} &= \frac{\delta r}{\delta t}. \\
        \lim_{u_x \rightarrow -\delta r/\delta t} c_s^{e+} &= 0,\\
        \lim_{u_x \rightarrow \delta r/\delta t} c_s^{e-} &= 0. \\
        \lim_{u_x \rightarrow -\delta r/\delta t} c_s^{e-} &= -\frac{\delta r}{\delta t}.
    \end{align}
\end{subequations}
Another point worth noting is that if one was to define the Mach number as the ratio of flow speed to the normal mode propagation speed, i.e.
\begin{subequations}
    \begin{align}
        {\rm Ma}^{+} &= \frac{u_x}{c_s^{+}-u_x},\label{eq:entropic:map}\\
        {\rm Ma}^{-} &= \frac{u_x}{c_s^{-}-u_x},\label{eq:entropic:man}
    \end{align}
\end{subequations}
the Mach number would evolve as shown in Fig.~\ref{Fig:entropic_mach_number}, which in effect means that the entropic equilibrium leads to a scheme that is stable in the limit of ${\rm Ma}\rightarrow\infty$.
Note that the deviation in the propagation speed of acoustic modes scales with the non-dimensional velocity with order three. This behavior is illustrated in Fig.~\ref{Fig:error_in_cs_entropic}.
\par Overall the following points could be made based on the Euler-level analyses:
\begin{itemize}
    \item The entropic equilibrium pressure tensor admits deviations from the Maxwell-Boltzmann pressure tensor that are not isotropic.
    \item Deviations are negligible for non-dimensional velocities as large as $0.5$ which is well above the domain of validity of the weakly compressible flow assumption.
    \item The deviations scale as $\propto {\left(\frac{u \delta t}{\delta r}\right)}^4$, see Fig.~\ref{Fig:entropic_eq_Pxx_Pyy}.
    \item The entropic equilibrium leads to a non-isotropic sound speed.
    \item In the limit of low non-dimensional flow velocities, this non-isotropic deviation scales out with $\propto {\left(\frac{u \delta t}{\delta r}\right)}^3$, see Fig.~\ref{Fig:error_in_cs_entropic}.
    \item The \emph{flow velocity-aware} nature of the sound speed allows for a system that adjusts its fastest propagating eigen-mode speed so as to always guarantee the condition of Eq.~\eqref{eq:CFL_strict_condition} is satisfied, see Fig.~\ref{Fig:sound_speed_entropic_vs_poly}.
    \item Defining the Mach number as the ratio of propagation speed of shear to normal modes, it is observed that entropic equilibrium can go to ${\rm Ma}^{\pm}\rightarrow\infty$ in the limit of $u_x \rightarrow \pm \delta r/\delta t$, see Fig.~\ref{Fig:entropic_mach_number}.
\end{itemize}
The slightly modified equilibrium pressure also affects dissipation at the Navier-Stokes level. The altered dissipation behavior is discussed in the next section.
\section{Navier-Stokes-level dynamics: dissipation\label{sec3}}
\subsection{1-D lattice: Bulk viscosity}
\begin{figure}[h!]
	\centering
	\includegraphics[width=5.7cm,keepaspectratio]{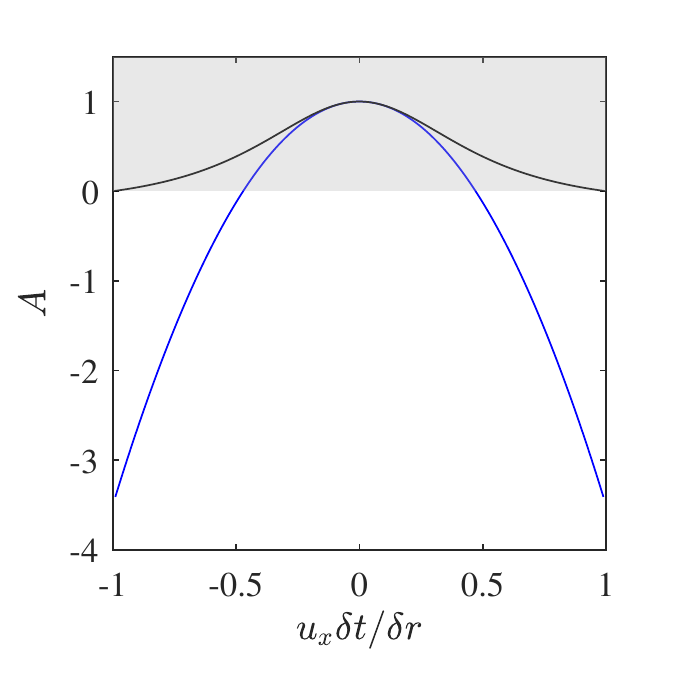}
	\caption{Comparison of the effective bulk viscosity: (blue) polynomial and (black) entropic equilibria.}
	\label{Fig:bulk_visc_entrop_vs_lbgk}
\end{figure}

\begin{figure}[h!]
	\centering
	\includegraphics[width=6cm,keepaspectratio]{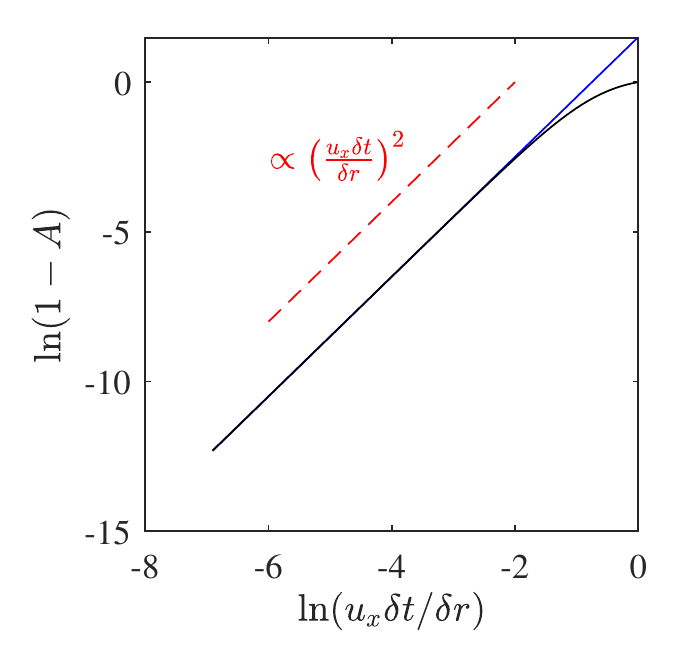}
	\caption{Scaling of the error in effective 1-D viscosity, i.e. $1-A$: (blue) polynomial and (black) entropic equilibria.}
	\label{Fig:bulk_viscosity_scaling}
\end{figure}

To illustrate the impact of the entropic equilibrium on dissipation let us first start with a 1-D system with three discrete velocities. In 1-D there is only acoustic modes, contrary to 2- and 3-D cases where shear modes are also present. Performing the classical Chapman--Enskog analysis, at order $\varepsilon^2$ the following momentum balance equation is recovered:
\begin{equation}
    \partial_t^{(2)}\rho u_x + \partial_x\left( 1-\beta\right)\Pi_{xx}^{(1)} = 0,
\end{equation}
where after some algebra the non-equilibrium stress tensor $\Pi_{xx}^{(1)}$ for any equilibrium pressure of the form:
\begin{equation}
    P_{xx} = \rho \varsigma^2 + P^{*}_{xx},
\end{equation}
results in:
\begin{equation}\label{eq:noneq_second_order_moment}
    \Pi_{xx}^{(1)} = -\frac{1}{2\beta} \left[2 A \rho \varsigma^2 \partial_x u_x + B \partial_x \rho\right],
\end{equation}
with
\begin{equation}\label{eq:entorpic_bulk}
    A = \left(1-\frac{3}{2}\frac{u_x^2}{\varsigma^2} - \frac{3}{2\rho \varsigma^2}u_x \partial_{u_x} P^{*}_{xx} - \frac{{(\partial_{u_x} P^{*}_{xx})}^2}{2\rho^2 \varsigma^2} - \frac{\partial_{\rho} P^{*}_{xx}}{2 \varsigma^2}\right),
\end{equation}
and
\begin{equation}\label{eq:entorpic_bulk_2}
    B= -3u_x\partial_{\rho} P^{*}_{xx} - \frac{\varsigma^2}{\rho}\partial_{u_x}P^{*}_{xx} - \frac{\partial_{u_x}P^{*}_{xx} \partial_{\rho} P^{*}_{xx}}{\rho} - u_x^3. 
\end{equation}
For the sake of clarity, in the remainder of the article, we will refer to \eqref{eq:entorpic_bulk} as the effective viscosity and \eqref{eq:entorpic_bulk_2} as the compressibility error at the Navier-Stokes level. For the case of the polynomial equilibria, i.e. both second-order and product form, one gets:
\begin{equation}\label{eq:LBGK_bulk}
    A = \left(1-\frac{3}{2}\frac{u_x^2}{\varsigma^2}\right),
\end{equation}
and:
\begin{equation}\label{eq:bulk_compressibility_error_polynomial}
    B = - u_x^3,
\end{equation}
which, neglecting the second term, indicates that the corresponding partial differential equation is only dissipative for:
\begin{equation}\label{eq:bulk_poly_positivity}
    \lvert u_x \lvert \leq \varsigma\sqrt{\frac{2}{3}}.
\end{equation}
Note that this maximum velocity is larger than the one imposed by CFL condition, i.e. Eq.~\eqref{eq:CFL_lmit_absolute_poly}. In the case of the entropic equilibrium:
\begin{equation}\label{eq:entropic_bulk}
    A = 1-\frac{3}{2}\frac{u_x^2}{\varsigma^2} + \frac{{(u_x/\varsigma)}^2+3\,{(u_x/\varsigma)}^4-2\,\sqrt{{(u_x/\varsigma)}^2+1}+2}{2{(u_x/\varsigma)}^2+2}
\end{equation}
and:
\begin{equation}\label{eq:bulk_compressibility_error_entropic}
    B = 0.
\end{equation}
Clearly for the entropic equilibrium in the limit of $u_x\rightarrow\delta r/\delta t$, the effective viscosity $A$ goes to zero.
The effective viscosities of Eqs.~\eqref{eq:LBGK_bulk} and \eqref{eq:entropic_bulk} are compared and shown in Fig.~\ref{Fig:bulk_visc_entrop_vs_lbgk}.
Furthermore, it is interesting to note that while the entropic equilibrium guarantees a positive effective viscosity it maintains the second-order convergence to the nominal viscosity, i.e. $A=1$, in the limit of $u_x\delta t/\delta r\rightarrow 0$, as illustrated in Fig.~\ref{Fig:bulk_viscosity_scaling}.
Going even further and comparing the compressibility errors, i.e. $B$, it is observed that the entropic equilibrium, contrary to polynomial equilibrium, does not have such errors. For the polynomial equilibria this error scales out at order three. \\
While two types of error were shown to exist in the Navier-Stokes level dissipation term, it is still unclear how each term would affect the dissipation of the eigen-modes derived in section \ref{sec2}. This issue will be discussed in the next section.
\subsection{Analysis of linearized Navier-Stokes equations in the limit of vanishing wave-number}
\begin{figure}[h!]
	\centering
	\includegraphics[width=9cm,keepaspectratio]{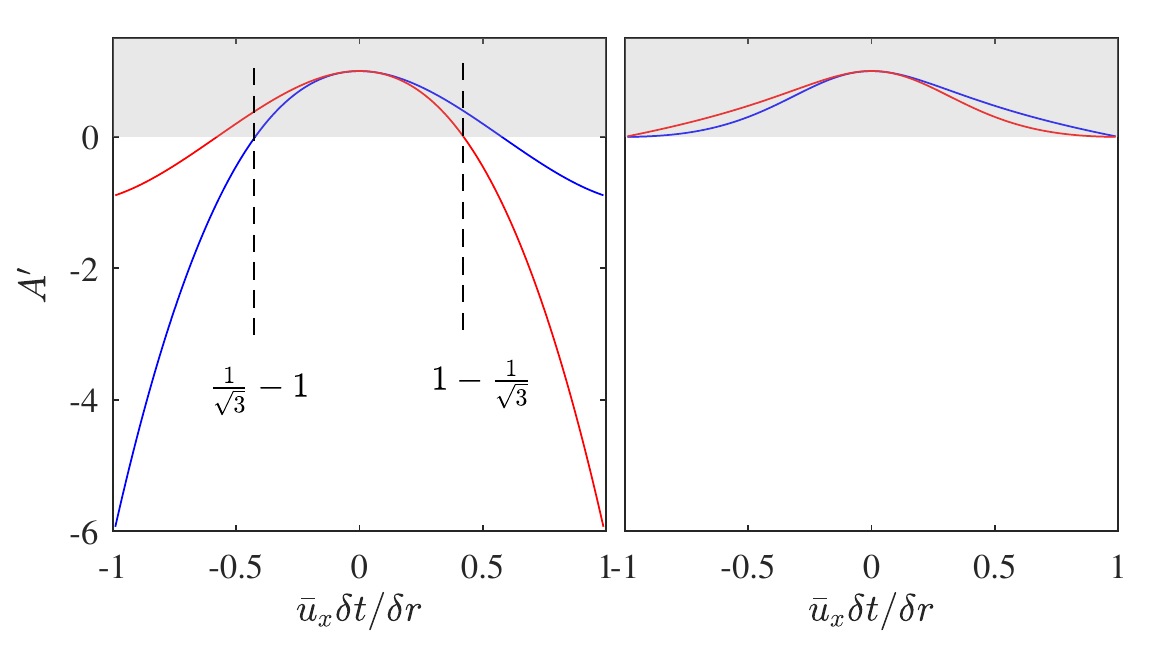}
	\caption{Overall dissipation rate of linearized hydrodynamic equations for (left) polynomial and (right) entropic equilibria. Red lines show dissipation of $c_s^+$ mode and blue lines that of $c_s^-$ mode.}
	\label{Fig:overall_bulk_viscosity}
\end{figure}

In the previous section we observed that two types of errors affecting the Navier-Stokes-level dissipation could be observed: One we called effective viscosity and another one we introduced as compressiblity error. While the role of the former is quite clear, even from its name, the \emph{real} effect of the latter on viscosity is rather abstract. To clarify the role of each error we will analyze the corresponding linearized equations in the limit of vanishing wave-number, i.e. $k_x\rightarrow 0$. For the sake of clarity we will first start with the simple case of polynomial equilibria. To linearize the system of partial differential equations at the Euler+Navier-Stokes level we separate velocity and density into a constant contribution and a  perturbation, 
\begin{subequations}
    \begin{align}
        \rho &= \bar{\rho} + \rho',\\
        u_x &= \bar{u}_x + u_x'.
    \end{align}
\end{subequations}
Introducing these into the system of partial differential equations and only keeping first-order perturbations we obtain the linearized system:
\begin{subequations}
    \begin{align}
        \partial_t \rho' &= -\bar{\rho} \partial_x u_x' - \bar{u}_x \partial_x \rho',\label{eq:lin_continuity_1d}\\
        \partial_t u_x' &= -\bar{u}_x\partial_x u_x' - \frac{\varsigma^2}{\bar{\rho}}\partial_x \rho' + 2\nu A \partial_x^2 u_x' + \frac{\nu B}{\bar{\rho}^2\varsigma^2} \partial_x^2 \rho',\label{eq:lin_NS_1d}
    \end{align}
\end{subequations}
where $B=B(\bar{\rho},\bar{u}_x)$ and $A=A(\bar{\rho},\bar{u}_x)$. Note the last term in Eq.~\eqref{eq:lin_NS_1d} stemming from the compressibility error. Next we consider the perturbations to be monochromatic plane waves, i.e.
\begin{subequations}
    \begin{align}
        \rho' &= \hat{\rho} \exp{\left(\mathrm{i}(\omega t - k_x x)\right)},\\
        u_x' &= \hat{u}_x \exp{\left(\mathrm{i}(\omega t - k_x x)\right)},
    \end{align}
\end{subequations}
where $\mathrm{i}=\sqrt{-1}$. Introducing the perturbations into Eqs.~\eqref{eq:lin_continuity_1d} and \eqref{eq:lin_NS_1d} we end up with the following system:
\begin{subequations}
    \begin{align}
    \omega \hat{\rho} &= \mathrm{i}\bar{\rho}k_x \hat{u}_x + \sqrt{-1} k_x \bar{u}_x \hat{\rho},\\
    \omega \hat{u}_x &= \mathrm{i} k_x \bar{u}_x \hat{u}_x + \frac{\varsigma^2}{\bar{\rho}} k_x \hat{\rho} - 2\nu A k_x^2 \hat{u}_x - \frac{\nu B}{\bar{\rho}^2\varsigma^2} k_x^2 \hat{\rho}.
    \end{align}
\end{subequations}
Finally, we solve the corresponding eigen-value problem and Taylor-expand corresponding solutions around $k_x=0$; Keeping terms of order $k_x$ and $k_x^2$ we get the following eigen-values:
\begin{subequations}
    \begin{align}
        \omega_{c_s^+} &= (\bar{u}_x + \varsigma) k_x + \mathrm{i}\nu \left(A + \frac{B}{2\bar{\rho}\varsigma^3}\right)k_x^2,\\
        \omega_{c_s^-} &= (\bar{u}_x - \varsigma) k_x + \mathrm{i}\nu \left(A - \frac{B}{2\bar{\rho}\varsigma^3}\right)k_x^2.
    \end{align}
\end{subequations}
Looking at the dissipation terms, i.e. $\propto k_x^2$, interesting observations on the role of each type of error could be made: (a) The existence of the compressibility error results in anisotropy in the dissipation of acoustic modes causing $c_s^+$ and $c_s^-$ to dissipate at different rates even though the dispersion is isotropic for polynomial equilibria. (b) The average of the dissipation rates of the two acoustic modes is modulated by the effective viscosity coefficient, $A$.\\
Extending this analysis to the more general form of the equilibrium as in previous sections the following eigen-values are recovered:
\begin{subequations}
    \begin{align}
    \omega_{c_s^+} &= c_s^+ k_x \nonumber\\ &+ \mathrm{i} \nu A\left(1 + \frac{\varsigma^2\partial_{\bar{u}_x}P_{xx}^* + B/A}{2\bar{\rho}\varsigma^2\sqrt{\varsigma^2+\partial_{\bar{\rho}}P_{xx}^*+(\partial_{\bar{u}_x}P_{xx}^*/2\bar{\rho})^2 }}\right)k_x^2,\\
    \omega_{c_s^-} &= c_s^- k_x \nonumber\\ &+  \mathrm{i} \nu A\left(1 - \frac{\varsigma^2 \partial_{\bar{u}_x}P_{xx}^* + B/A}{2\bar{\rho}\sqrt{\varsigma^2+\partial_{\bar{\rho}}P_{xx}^*+(\partial_{\bar{u}_x}P_{xx}^*/2\bar{\rho})^2 }}\right)k_x^2,
    \end{align}
\end{subequations}
where $c_s^+$ and $c_s^-$ are those derived in Eqs.~\eqref{eq:cs_plus_gen} and \eqref{eq:cs_minus_gen} and $P_{xx}^*=P_{xx}^*(\bar{\rho},\bar{u}_x)$. Plugging in the entropic equilibrium:
\begin{subequations}
    \begin{align}
    \omega_{c_s^+} &= c_s^{e+} k_x \nonumber\\ &+ \mathrm{i} \nu A\left(1 + \frac{1-\varsigma \bar{u}_x\left(\sqrt{{\left(\frac{\bar{u}_x}{\varsigma}\right)}^2+1}-1\right)}{\sqrt{2\sqrt{{\left(\frac{\bar{u}_x}{\varsigma}\right)}^2+1}-1}}\right)k_x^2,\label{eq:omega_csp_entropic}\\
    \omega_{c_s^-} &= c_s^{e-} k_x \nonumber\\ &+  \mathrm{i} \nu A\left(1 - \frac{1-\varsigma \bar{u}_x\left(\sqrt{{\left(\frac{\bar{u}_x}{\varsigma}\right)}^2+1}-1\right)}{\sqrt{2\sqrt{{\left(\frac{\bar{u}_x}{\varsigma}\right)}^2+1}-1}}\right)k_x^2\label{eq:omega_csm_entropic},
    \end{align}
\end{subequations}
where $c_s^{e+}$ and $c_s^{e-}$ were derived in Eqs.~\eqref{eq:cs_plus_ent} and \eqref{eq:cs_minus_ent}. The interesting point worth noting here is that even though for the entropic equilibrium it was shown that $B=0$, writing the equations in terms of the eigen-modes of the system one recovers a compressibility-like error term.\\
Now that we have a closed-form solution for the total dissipation rate of the linearized hydrodynamic equations, noted as $A'$, defined in Eqs.~\eqref{eq:omega_csp_entropic} and \eqref{eq:omega_csm_entropic} as:
\begin{equation}
    \omega_{c_s} = c_s^e k_x + \mathrm{i}\nu A{'}k_x^2,
\end{equation}
well-posedeness of the corresponding wave system can be assessed by looking at the sign of the dissipation rates. The results are shown in Fig.~\ref{Fig:overall_bulk_viscosity}.
It is observed that the entropic equilibrium always remains dissipating while the polynomial equilibrium keeps that property within a certain range of velocity:
\begin{equation}
    \frac{1}{\sqrt{3}}-1\leq \frac{u_x\delta t}{\delta r} \leq 1 - \frac{1}{\sqrt{3}},
\end{equation}
which interestingly enough matches exactly the CFL condition on eigen-modes as shown in Fig.~\ref{Fig:sound_speed_entropic_vs_poly}.\\
The next section will discuss the dissipation rates of shear modes.
\subsection{Multi-dimensional lattices: Shear viscosity}
\begin{figure}[h!]
	\centering
	\includegraphics[width=6.5cm,keepaspectratio]{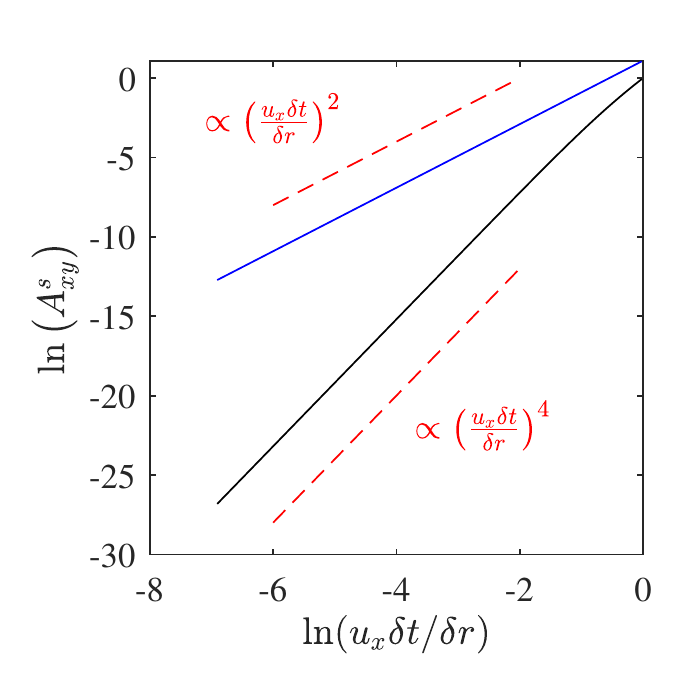}
	\caption{Comparison of scaling of the effective viscosity in shear components of the viscous stress tensor for the (black) entropic and (blue) second-order polynomial equilibria.}
	\label{Fig:error_scaling_shear}
\end{figure}

\begin{figure}[h!]
	\centering
	\includegraphics[width=6.5cm,keepaspectratio]{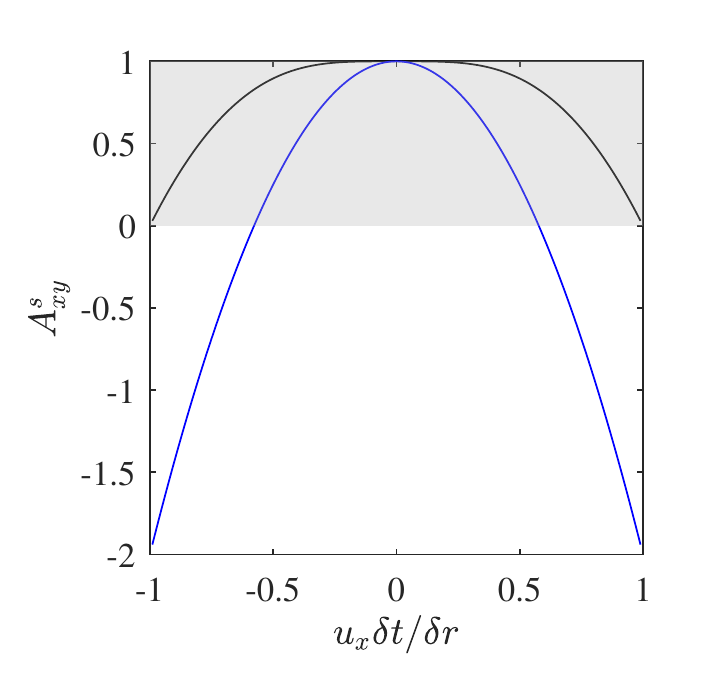}
	\caption{Comparison of positivity domain of the effective viscosity in shear components of the viscous stress tensor for the (black) entropic and (blue) second-order polynomial equilibria.}
	\label{Fig:error_positive_shear}
\end{figure}

The multiscale analysis of the previous section can be extended to multiple dimensions, detailed in \ref{app:CE_multiD}. Decomposing the non-equilibrium stress tensor into two contributions as:
\begin{equation}
    \Pi_2^{(1)} = -\frac{1}{2\beta}\left(\bm{S}^* + \bm{D}^*\right),
\end{equation}
where:
\begin{multline}
    \bm{S}^* = {\left[(\rho\varsigma^2\bm{I}+\bm{P}^*)\cdot\bm{\nabla}\bm{u}\right]}+{\left[(\rho\varsigma^2\bm{I}+\bm{P}^*)\cdot\bm{\nabla}\bm{u}\right]}^\dagger + \bm{E}^s.
\end{multline}
Here $\bm{E}^s$ regroups errors specific to the second-order polynomial, detailed in Eq.~\eqref{eq:2nd_order_poly_err_multiD}. Focusing on off-diagonal components and following the presentation of the previous section, let us introduce an effective viscosity, $\bm{A}^s$ such that:
\begin{equation}
    S^*_{\alpha\beta} + E_{\alpha\beta}^s = \rho \varsigma^2 \left[A_{\alpha\beta}^s \partial_\alpha u_\beta + A_{\beta\alpha}^s \partial_\beta u_\alpha\right] + B_{\alpha}^s\partial_\alpha \rho + B_{\beta}^s\partial_\beta \rho.
\end{equation}
In the case of the product form equilibrium:
\begin{equation}
    A_{\alpha\beta}^s = A_{\beta\alpha}^s = 1,
\end{equation}
while for the second order polynomial:
\begin{equation}
    A_{\alpha\beta}^s = 1 - \frac{u_\alpha}{\varsigma^2}(u_\alpha+2u_\beta),
\end{equation}
and the entropic equilibrium, Eq.~\eqref{eq:Entropic_isothermal_EDF}:
\begin{equation}
    A_{\alpha\beta}^s = 1 + \frac{P_{\alpha\alpha}^*}{\rho\varsigma^2} = 2\sqrt{{\left(\frac{u_\alpha}{\varsigma}\right)}^2 + 1} - {\left(\frac{u_\alpha}{\varsigma}\right)}^2 - 1.
\end{equation}
The dependence of $A_{xy}^s$ on $u_x\delta t/\delta r$ with $u_y=0$ for all equilibria are illustrated in Fig.~\ref{Fig:error_scaling_shear}.
The deviations in the entropic equilibrium are shown to scale out at order four while for the second order polynomial equilibrium the order is only two. Furthermore, the effective viscosity $A_{xy}^s$ for the entropic equilibrium is positive definite for $\lvert u_x\lvert\in [0,\delta r/\delta t]$, while for the second order polynomial equilibrium it is only positive for:
\begin{equation}\label{eq:shear_postive_2nd_polynomial}
    \lvert u_x\lvert \leq\varsigma.
\end{equation}
This point is illustrated in Fig.~\ref{Fig:error_positive_shear}.
\par Based on the analysis of the Navier-Stokes level equations the following observations could be made:
\begin{itemize}
    \item Contrary to polynomial equilibria, the Navier-Stokes-level viscosity of normal modes recovered by the entropic discrete equilibrium is positive definite, for $-\delta r/\delta t\leq u_x \leq \delta r/\delta t$, see Fig.~\ref{Fig:bulk_visc_entrop_vs_lbgk}. Polynomial equilibria lose positivity following Eq.~\eqref{eq:bulk_poly_positivity}.
    \item The deviation in the viscosity tied to compressibility, i.e. $\propto \partial_x \rho$, for the polynomial equilibria scales with $\propto {\left(\frac{u_x\delta t}{\delta r}\right)}^3$, see Eq.~\eqref{eq:bulk_compressibility_error_polynomial}. The entropic equilibrium on the other hand does not admit such errors, see Eq.~\eqref{eq:bulk_compressibility_error_entropic}.
    \item An analysis of the dissipation of eigen modes of different models reveals that entropic normal eigen modes are indeed subject to a compressibility-like error, see Eqs.~\eqref{eq:omega_csp_entropic} and \eqref{eq:omega_csm_entropic}.
    \item Compressibility errors cause the two acoustic modes $c_s^+$ and $c_s^-$ to dissipate at different rates, even in the limit of $k_x\rightarrow 0$.
    \item Following the trend set in previous sections, the polynomial equilibria result in an overall dissipation rate that is only positive for both modes in $\lvert u_x\lvert \in [0\text{ }\frac{\delta r}{\delta t}-\varsigma]$.  For the entropic equilibrium on the other hand the overall dissipation rate of the linearized hydrodynamic equations is positive definite for all modes in $\lvert u_x\lvert \in [0\text{ }\delta r/\delta t]$.
    \item Looking at Fig.~\ref{Fig:overall_bulk_viscosity}, it is seen that the anisotropy in dissipation rate of the two acoustic modes is much less pronounced for the entropic equilibrium.
    \item As for normal modes, the Navier-Stokes level dissipation rate of shear modes in the entropic equilibrium is positive definite. The second-order polynomial equilibrium on the other hand is only positive under the condition of Eq.~\eqref{eq:shear_postive_2nd_polynomial}. Note that for a velocity vector with non-zero components in the $y-$ and $z-$directions this condition becomes more restrictive.
    \item The deviations in the shear dissipation rate scale out with $\propto {\left(\frac{u_\alpha\delta t}{\delta r}\right)}^4$ while for the second order polynomial equilibrium they scale as $\propto {\left(\frac{u_\alpha\delta t}{\delta r}\right)}^2$.
\end{itemize}
\section{Numerical applications\label{sec4}}
\subsection{Linear stability analysis}
\subsubsection{von Neumann spectral analysis: introduction}
In the context of the von Neumann linear analysis, the discrete time-evolution system of equations is expanded around a reference state $\bar{f}_i\left(\bar{\rho},\bar{u}\right)$ via a first-order Taylor expansion:
\begin{equation}
f_i \approx \bar{f}_i + f^{'}_i,
\end{equation}
where $f^{'}_i$ is the linear perturbation. Defining the discrete collision operator as:
\begin{equation}
    \Omega_i = 2\beta\left(f_i^{\rm eq} - f_i\right),
\end{equation}
the linearized operator is:
\begin{equation}
\Omega_i(f_i) \approx \Omega_i\rvert_{\bar{f}_i} + \mathcal{J}_{ij}f^{'}_{j},
\end{equation}
where $\mathcal{J}_{ij}$ is the Jacobian of the collision operator evaluated about $\bar{f}_{j}$, i.e
\begin{equation}
    \mathcal{J}_{ij}=\partial_{f_{j}}\Omega_{i}\rvert_{\bar{f}_{j}}.
\end{equation}
Placing back these expressions into the discrete time-evolution equations one obtains~\cite{worthing_stability_1997}:
\begin{equation}
f_i^{'}\left( \bm{r}+\bm{c}_i \delta t,t+\delta t\right) = \left(\delta_{ij}+\mathcal{J}_{ij}\right) f_{j}^{'}\left( \bm{r},t\right).
\label{eq:PerturbedDiscreteLBE}
\end{equation}
Detailed expression for the Jacobian of the entropic equilibrium is given in Appendix~\ref{app_jacobian}. The last step of the von Neumann analysis is to assume that perturbations $f'_i$ are monochromatic plane waves of the form:
\begin{equation}
    f'_i = F_i\exp{[\mathrm{i}(\bm{k}\cdot\bm{r}-\omega_i t)]},
\end{equation}
where $F_i$ is the wave amplitude, $\mathrm{i}$ is the imaginary unit, $\vert\vert\bm{k}\vert\vert=k$ is the wave-number, and $\omega$ is the complex time frequency of the wave. $k$ is related to the wave-length of $f'_i$, whereas $\Im(\omega)$ and $\Re(\omega)$ are related to its attenuation and propagation speed. Introducing these perturbations into Eq.~\eqref{eq:PerturbedDiscreteLBE} one obtains the following eigenvalue problem of size $Q$ :
\begin{equation}\label{eq:VN_eigen}
\bm{M F} = \exp{(-\mathrm{i}\omega_i)} \bm{F},
\end{equation}
where $\bm{F}$ is the eigenvector composed of all amplitudes. $\bm{M}$ is the matrix associated to Eq.~\eqref{eq:PerturbedDiscreteLBE}. This matrix can be expressed as :
\begin{equation}
\bm{M} =\bm{E}\left[ \bm{\delta} +\mathcal{J}\right], 
\end{equation}
with
\begin{equation}
E_{ij} = \exp[-\mathrm{i}(\delta t\bm{c}_i\cdot\bm{k})]\delta_{ij}.
\end{equation}
Note that the matrix $\bm{M}$ and the eigenvalue problem~\eqref{eq:VN_eigen} depend on the mean flow ($\bar{\rho},\lvert\lvert \bar{\bm{u}}\lvert\lvert$), the wave-number ($k_x$ and $k_y$) and the relaxation frequency $\beta$. This means that for each set of these parameters the eigenvalue problem needs to be solved to obtain the corresponding values of $\Re{(\omega)}$ and $\Im{(\omega)}$. Doing so, the spectral properties can be obtained.
\subsubsection{Linear stability domain}
\begin{figure}[h!]
	\centering\includegraphics[width=6cm,keepaspectratio]{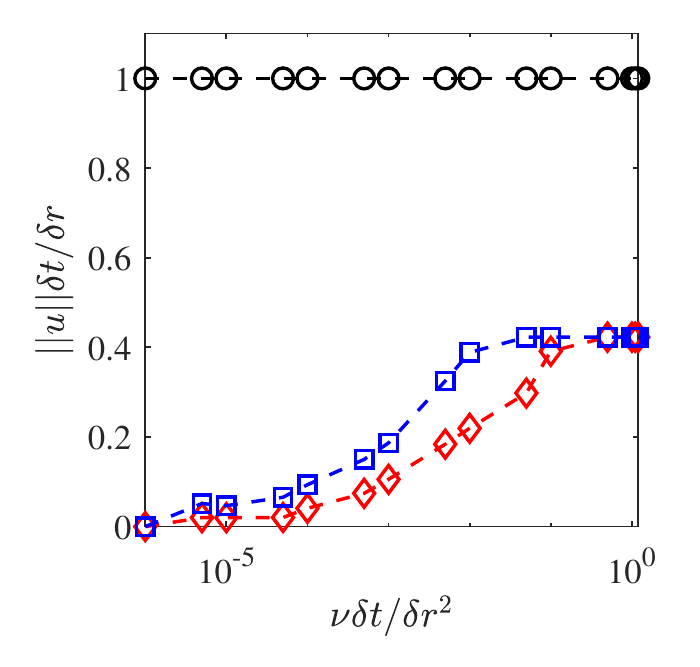}
	\caption{Comparison of linear stability domain of different equilibrium distribution functions: (red with diamond markers) second order polynomial, (blue with square markers) product form and (black with circular markers) entropic. }
	\label{Fig:stability_all}
\end{figure}

\begin{figure}[h!]
	\centering
	\includegraphics[width=6cm,keepaspectratio]{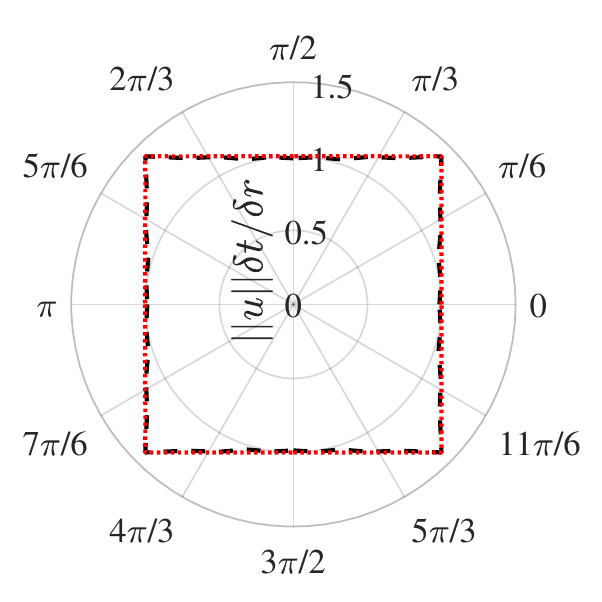}
	\caption{Directional stability domain of the entropic equilibrium for two different non-dimensional viscosities: (dashed black) $10^{-5} $and (dotted red) $0.1$.}
	\label{Fig:stability_entropic}
\end{figure}

\begin{figure}[h!]
	\centering
	\includegraphics[width=6cm,keepaspectratio]{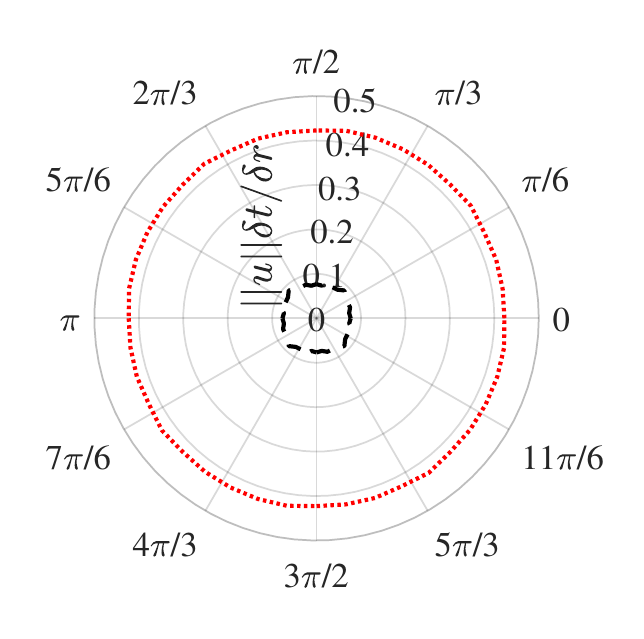}
	\caption{Directional stability domain of the polynomial equilibrium for two different non-dimensional viscosities: (dashed black) $10^{-5} $and (dotted red) $0.1$.}
	\label{Fig:stability_eq2}
\end{figure}

\begin{figure}[h!]
	\centering
	\includegraphics[width=6cm,keepaspectratio]{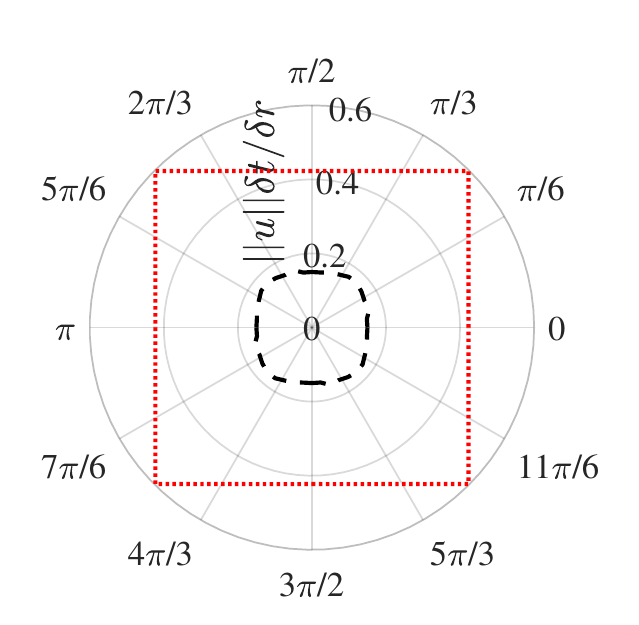}
	\caption{Directional stability domain of the product form equilibrium for two different non-dimensional viscosities: (dashed black) $10^{-5} $and (dotted red) $0.1$.}
	\label{Fig:stability_eq4}
\end{figure}

To assess the linear stability domain via the von Neumann approach, for a given set of properties $(\bar{\rho},\lvert\lvert \bar{\bm{u}}\lvert\lvert,\nu\delta t/\delta r^2)$ the eigen-value problem of Eq.~\eqref{eq:VN_eigen} is solved for $k_x\in[0\text{ }\pi]$ and $k_y\in[0\text{  }\pi]$ and $\theta=\cos^{-1}\left(\frac{u_x}{\lvert\lvert \bm{u}\lvert\lvert}\right) \in [0\text{  }\pi/2]$. The solver is considered to be linear-stable for the set of parameters if it guarantees negative attenuation rate for all modes and values of $k_x,k_y$ and $\theta$. We first quantify the stability domain of the discrete solvers by finding the maximum non-dimensional velocity for which linear stability is guaranteed for a range of non-dimensional viscosities. Here we consider $\nu\delta t/\delta r^2\in\{10^{-6},5\times 10^{-6},10^{-5},5\times 10^{-5},10^{-4},5\times 10^{-4}, 10^{-3},5\times 10^{-3},10^{-2},5\times 10^{-2},0.1,0.5,1,1.1,1.2\}$. The stability domains of three different equilibria, namely entropic, second order polynomial and product form are shown in Fig.~\ref{Fig:stability_all}.
It can be noted that the entropic equilibrium had, by far, the widest domain of stability guaranteeing unconditional stability which is unaffected by the value of the non-dimensional viscosity~\cite{hosseini_development_2020}. For the second-order polynomial and product form equilibria on the other hand, in the limit of vanishing non-dimensional viscosities the solvers become unconditionally unstable. On the other end of the spectrum, i.e. large non-dimensional viscosities the maximum speed encounters a threshold which is exactly the one shown in Eq.~\eqref{eq:CFL_lmit_absolute_poly}.\\
To better illustrate the stability behavior of the different equilibria, and especially to better show the unconditional linear stability of the entropic equilibrium, directional (orientation of velocity vector) stability domains are shown in Figs.~\ref{Fig:stability_entropic}, \ref{Fig:stability_eq2} and \ref{Fig:stability_eq4}.
For the entropic equilibrium, the solver linear stability covers the entire lattice and confirms observations from characteristics analysis, i.e. unconditional stability.
For the second-order polynomial for large non-dimensional viscosities the linear stability domains is close to isotropic and limited by the condition of Eq.~\eqref{eq:CFL_lmit_absolute_poly}. 
The product form equilibrium exhibits a directional behavior similar to the entropic one for large non-dimensional viscosities, however with a much smaller tolerated maximum velocity. The largest reachable non-dimensional velocity is reached when the velocity vector is oriented in the direction of one of the diagonal links of the lattice,
\begin{equation}
    \lvert\lvert \bm{u} \lvert\lvert=\sqrt{2}\left( \frac{ \delta r}{\delta t} - \varsigma\right) = 0.598 \frac{ \delta r}{\delta t}.
\end{equation}
\subsubsection{Spectral dispersion/dissipation}
\begin{figure}[h!]
	\centering
	\includegraphics[width=8cm,keepaspectratio]{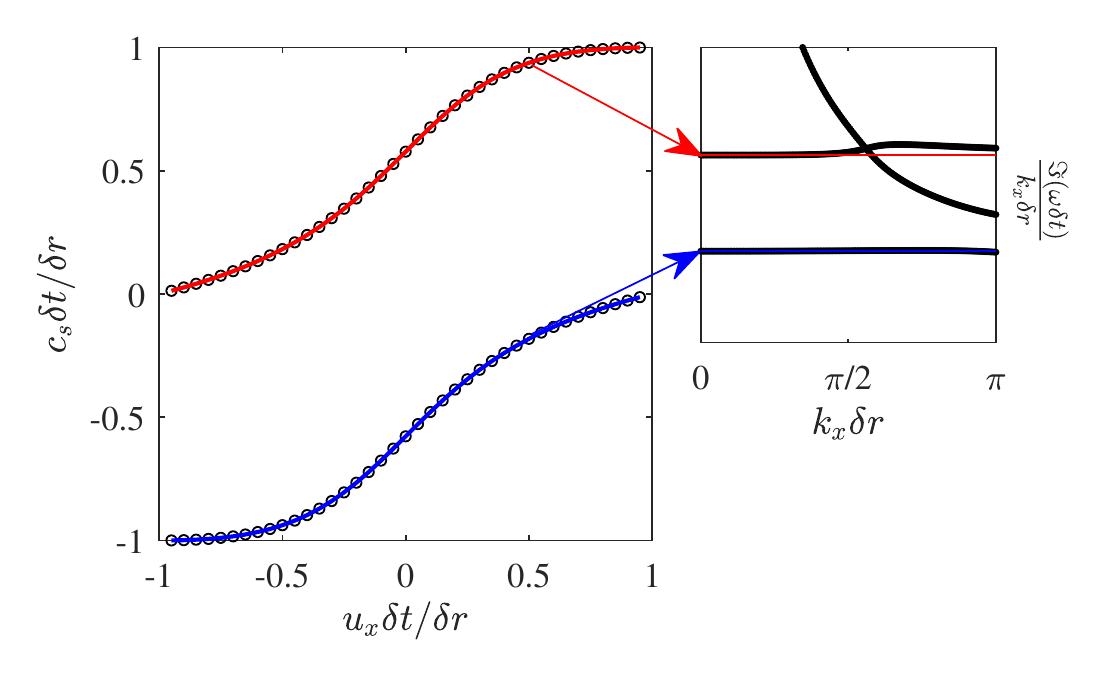}
	\caption{Speed of normal modes for the entropic equilibrium as a function of velocity, (black markers) as obtained from the spectral analysis of the full discrete system and (red line) from the characteristics analysis, i.e. Eqs.~\eqref{eq:cs_plus_ent} and \eqref{eq:cs_minus_ent}. The insert on the right hand side shows (black) the typical discrete spectrum. Red lines show dissipation of $c_s^+$ mode and blue lines that of $c_s^-$ mode. }
	\label{Fig:spectral_sound_speed}
\end{figure}

\begin{figure}[h!]
	\centering
	\includegraphics[width=8cm,keepaspectratio]{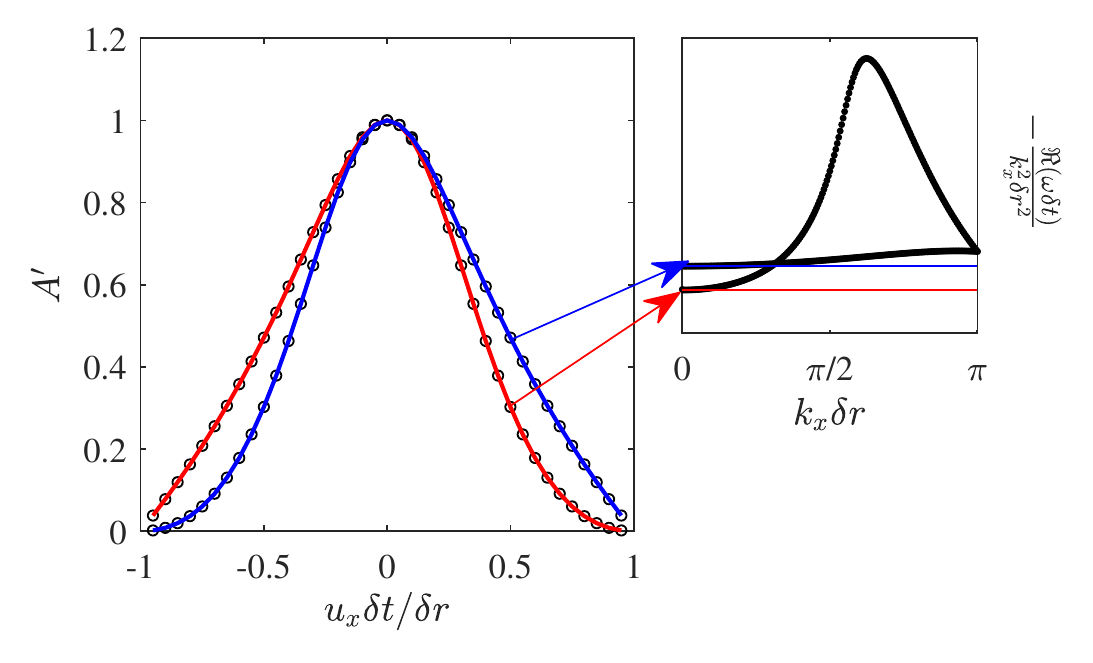}
	\caption{Effective overall viscosity, $A'$, for the entropic equilibrium as a function of velocity, (black markers) as obtained from the spectral analysis of the full discrete system and (red line) from the corresponding Navier-Stokes linearized equations, i.e. Eqs.~\eqref{eq:omega_csp_entropic} and \eqref{eq:omega_csm_entropic}. The insert on the right hand side shows (black) the typical discrete spectrum. Red lines show dissipation of $c_s^+$ mode and blue lines that of $c_s^-$ mode.}
	\label{Fig:spectral_bulk_dissipation}
\end{figure}

In this section we mainly aim at confirming two observations made via theoretical analyses of the hydrodynamic equations: (a) local-velocity awareness of the entropic sound speed and (b) the effect of different types of error on the effective normal dissipation rate. To that end we will compare results from von Neumann analysis of the full discrete system to previously obtained analytical expressions.\\
First we look into the dispersion of normal modes. The spectral propagation speed is computed as $\Im(\omega)/k_x$. The calculations have been carried out for velocities $u_x\delta r/\delta t \in [-1\,\,1]$ and are compared to hydrodynamic limit predictions from the characteristics analysis. The results are shown in Fig.~\ref{Fig:spectral_sound_speed}.
The comparison shows that the analytical expressions derived in Eqs.~\eqref{eq:cs_plus_ent} and \eqref{eq:cs_minus_ent} exactly match the numerical results from spectral analysis at $k_x\rightarrow 0$.\\
The second parameter to be validate in this section is the effective dissipation rates of normal modes. The dissipation rate is extracted as $-\Re(\omega)\delta t/k_x^2 \delta r^2$ from the full spectral analysis and compared to analytical expressions derived in Eqs.~\eqref{eq:omega_csp_entropic} and \eqref{eq:omega_csm_entropic}. The calculations have been carried out for velocities $u_x\delta r/\delta t \in [-1\,\,1]$ and are compared to hydrodynamic limit predictions in Fig.~\ref{Fig:spectral_bulk_dissipation}.
The comparisons show exact agreement and confirm the analytical results derived in previous sections.
\subsection{Correction of leading-order bulk viscosity error}
\subsubsection{Rescaling bulk viscosity}
\begin{figure}[h!]
	\centering
	\includegraphics[width=6cm,keepaspectratio]{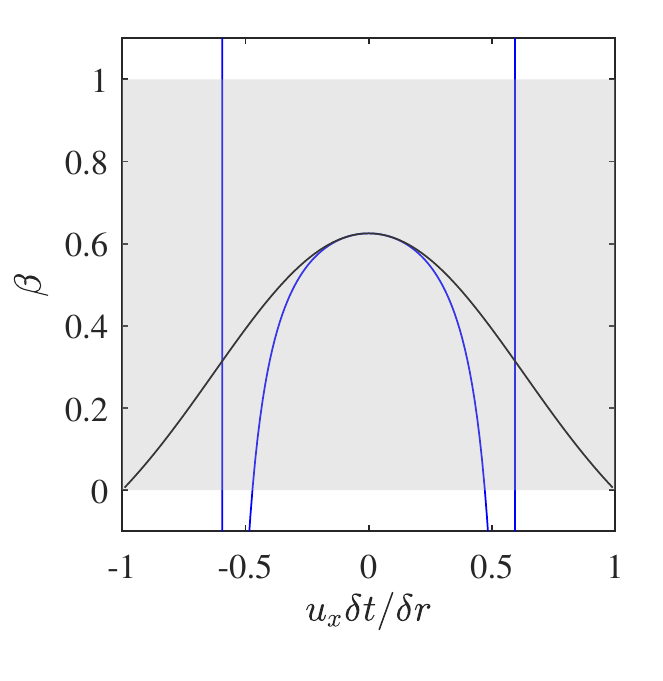}
	\caption{Variation in the redefined relaxation rate, Eq.~\eqref{eq:redefined_relaxation rate} as a function of local velocity: (blue) polynomial equilibrium and (black) entropic equilibrium. Here non-dimensional viscosity is taken to be $\nu\delta t/\delta r^2=0.1$.}
	\label{Fig:redefined_relaxation_rate}
\end{figure}

In the previous section, it was shown that at the Navier-Stokes level the bulk viscosity had two Galilean-variant errors, see Eq.~\eqref{eq:entorpic_bulk} and \eqref{eq:entorpic_bulk_2}. The first term, i.e. Eq.~\eqref{eq:entorpic_bulk} can be corrected by redefining the relaxation frequency as:
\begin{equation}\label{eq:redefined_relaxation rate}
    \beta = \frac{\delta t}{\frac{2\nu}{A \varsigma^2} + \delta t},
\end{equation}
which, for instance, for polynomial equilibria reduces to:
\begin{equation}\label{eq:redefined_relaxation rate_poly}
    \beta = \frac{\delta t}{\frac{2\nu}{\left(\varsigma^2-3u_x^2/2\right)} + \delta t}.
\end{equation}
In doing so, and before checking the effect of this correction two points have to be considered: (a) The effect of this redefinition of the relaxation rate on the second error term, i.e. Eq.~\eqref{eq:entorpic_bulk_2}, and (b) changes in the now-velocity-dependent and variable relaxation rate.
The changes induced in the relaxation rate by this correction are illustrated in Fig.~\ref{Fig:redefined_relaxation_rate}.
The key observation here is that for the polynomial equilibria the redefined relaxation rate grows much faster than for the entropic equilibrium, and that $\beta\in[0\text{ }1]$ is not always guaranteed. For this condition to be satisfied one must have:
\begin{equation}\label{eq:positivity_visc_A}
    \frac{2\nu}{\varsigma^2 A}\geq 0,
\end{equation}
which boils down to a condition of positivity of $A$ since both $\nu$ and $\varsigma$ are positive. As shown previously in Fig.~\ref{Fig:bulk_visc_entrop_vs_lbgk} this condition is not always satisfied for the polynomial and product form equilibria. For the entropic equilibrium on the other hand, because $A$ is positive definite, the redefined relaxation rate satisfies $\beta\in[0\text{ }1]$ for $\lvert u_x\lvert < \delta r/\delta t$.\\
The effect of this correction will be checked via numerical simulations in the next section.
\subsubsection{Validation via dissipation of isothermal pressure waves}
\begin{figure}[h!]
	\centering
	\includegraphics[width=6cm,keepaspectratio]{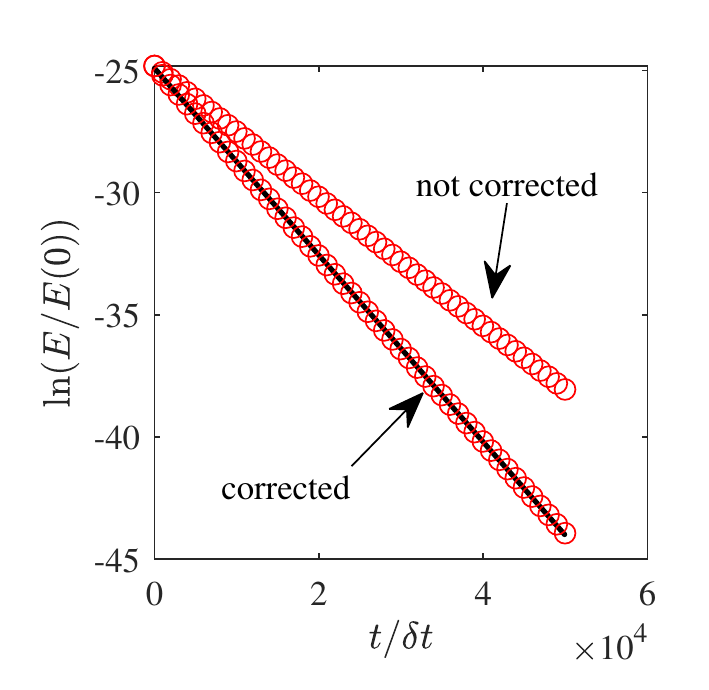}
	\caption{Evolution of total energy as a function of time as obtained from two simulations using the entropic equilibrium, one with the regular definition of $\beta$ from Eq.~\eqref{eq:relaxation_frequency} and one with the redefined $\beta$ from Eq.~\eqref{eq:redefined_relaxation rate}. $\nu\delta t/\delta r^2$ is set to $0.05$.}
	\label{Fig:measured_bulk_sim_example}
\end{figure}

In the limit of small velocities and density variations, the contributions from non-linear terms in the Navier-Stokes equation can be ignored and acoustic waves can be modeled through the linear theory including losses, which reads~\cite{kinsler_fundamentals_2000}:
\begin{equation}
	\partial_t^2 u_x = c_s^2\partial_x^2 u_x + \left( \frac{D+1}{D}\nu + \frac{\eta}{\rho} \right)\partial_t\partial_x^2 u_x,
\end{equation}
where the bulk viscosity $\eta$ in the context of the classical LB model is fixed at $\frac{D-1}{D}\rho\nu$. The exact solution of this equation can be shown to be of the form $u_x \propto \exp\left({\sqrt{-1} k_x x + \sigma t}\right)$ where:
\begin{equation}
	\sigma = -\left( \frac{2}{3}\nu + \frac{1}{2}\eta \right)k_x^2+\sqrt{-1} k_x c_s\sqrt{1-{\left( \frac{D+1}{D}\nu+\frac{\eta}{\rho} \right)}^2\frac{k_x^2}{c_s^2}}.
\end{equation}
The total energy, here defined as $E=\int_x \frac{1}{2}u^2+c_s^2\left( \rho - \rho_0 \right) dx$ will therefore decay in time as:
\begin{equation}\label{eq:energy_decay}
    E(t) = E(0) \exp{\left[-\left(\frac{D+1}{D}\nu+\frac{\eta}{\rho}\right)\frac{t}{k_x^2}\right]},
\end{equation}
with $k_x=\frac{2\pi}{L}$, where $L$ is the wavelength. To evaluate the effective bulk viscosity of the considered numerical schemes, we initialize a wave-function in a domain of size $N_x$ with periodic boundary conditions as:
\begin{equation}
	\rho = \rho_0 + \delta \rho \sin \left( \frac{2\pi x}{N_x} \right),
\end{equation}
for different values of initial velocity $U_0\in[-0.3\text{ }0.3]$. The waves are then left to evolve, and energy is stored for each time-step. The resulting time-evolution of the energy is then fitted with a function of the form of \eqref{eq:energy_decay} to extract the effective viscosity. This process is illustrated in Fig.~\ref{Fig:measured_bulk_sim_example}, with results obtained from a sample simulation.
Simulations were ran for different convection speeds for both entropic and product form equilibria both with and without correction of the relaxation frequencies. For the entropic equilibrium the corrected relaxation frequencies were computed according to Eq.~\eqref{eq:redefined_relaxation rate} while for the product form Eq.~\eqref{eq:redefined_relaxation rate_poly} was used.
\begin{figure}[h!]
	\centering
	\includegraphics[width=10cm,keepaspectratio]{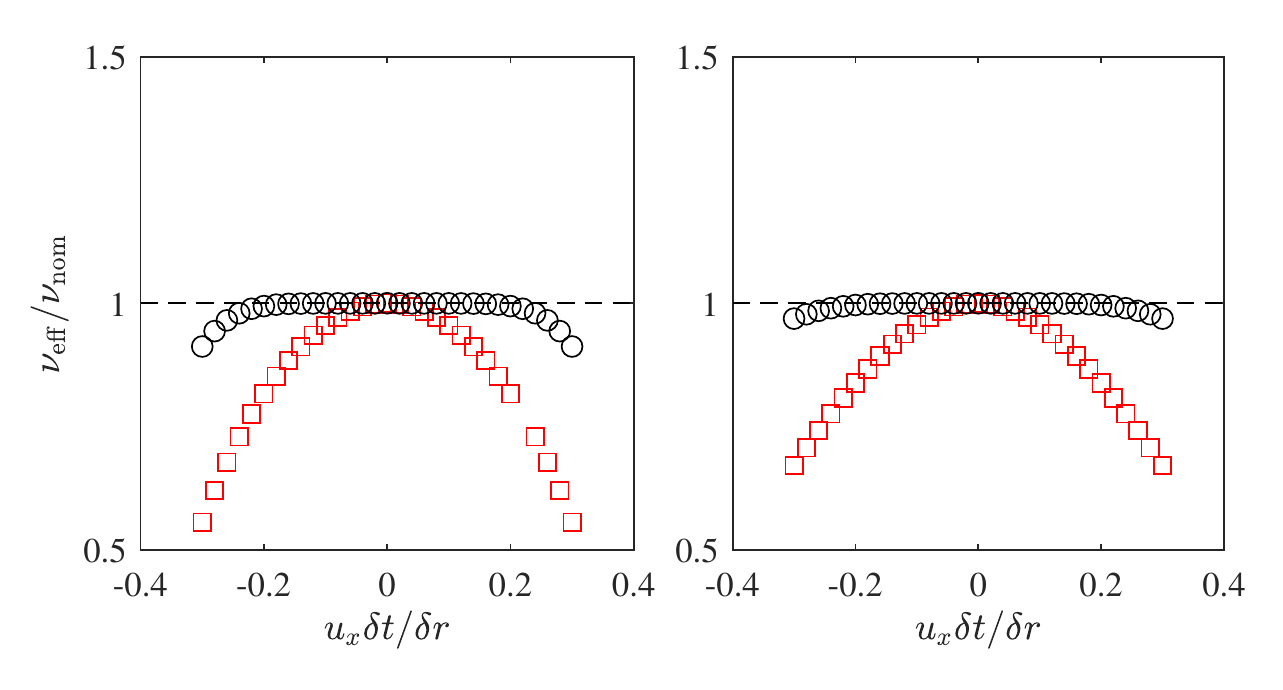}
	\caption{Effective viscosity as a function of non-dimensional velocity as obtained from the 1-D acoustic wave dissipation case for (left) product form and (right) entropic equilibria. Red square symbols show results with $\beta$  defined as in Eq.~\eqref{eq:relaxation_frequency} and black circular symbols results with $\beta$ as defined in Eq.~\eqref{eq:redefined_relaxation rate}. }
	\label{Fig:measured_bulk_sim}
\end{figure}
The obtained results are illustrated in Fig.~\ref{Fig:measured_bulk_sim}. Based on these results a number of observations can be made; The first one is that in the non-corrected forms, the product form equilibrium induces more deviations into the effective viscosity as compared to the entropic equilibrium. Second, the rescaling of the relaxation frequency does reduce the error in the effective  viscosity, however as expected it does not fully eliminate it.\\
\par Finally the outcomes of this final section can be summarized as follows:
\begin{itemize}
    \item Linear spectral analysis confirmed that the entropic equilibrium guarantees unconditional stability for any wave-number $\bm{k}$ for $\lvert u_\alpha\lvert\leq\delta r/\delta t$, see Fig.~\ref{Fig:stability_entropic}. The family of polynomial equilibria, both second order and product form have a more limited linear stability domain and are affected by the choice of the non-dimensional viscosity, see Fig.~\ref{Fig:stability_eq2}, \ref{Fig:stability_eq4} and \ref{Fig:stability_all}.
    \item The errors in effective viscosity, in 1-D, from Eq.~\eqref{eq:entorpic_bulk} can be, to a great extend, reduced via a simple rescaling of the relaxation frequency as given by Eq.~\eqref{eq:redefined_relaxation rate}. This is confirmed by Fig.~\ref{Fig:measured_bulk_sim}.
    \item The rescaling strategy applied to the entropic equilibrium has a number of advantages over its product form counter part: For the product form it can result in $\beta\notin [0\text{ }1]$, see Eq.~\eqref{eq:positivity_visc_A}. The corrected entropic model exhibits much less pronounced deviations as compared to the product form.
\end{itemize}
\section{Conclusions and discussion\label{sec5}}
Construction of discrete distribution functions in reduced kinetic models, whether it be moments-based methods such as the Grad system or discrete velocity methods such as the lattice Boltzmann method is crucial to both accuracy and stability of the system. While the classical approach, used for instance for classical polynomial equilibria in the lattice Boltzmann, consists in matching moments of interest in the equilibrium to their continuous counter-parts. The entropic construction provides an approach that satisfies target moments constraints within the range of operation and asymptotically vanishing sound speed which, as demonstrated in details here, leads to unconditional linear stability. 
Our detailed analysis of the entropic construction further showed that this asymptotic regularization, in the case of the entropic equilibrium, also leads to positive-definite Navier-Stokes-level dissipation rates. A simple local approach was also introduced to further reduce deviation in the effective viscosity of the model. The correction was also shown to maintain positive-definiteness only for the entropic equilibrium.

\begin{acknowledgments}
This work was supported by European Research Council (ERC) Advanced Grant  834763-PonD. 
Computational resources at the Swiss National  Super  Computing  Center  CSCS  were  provided  under the grant  s1066.
\end{acknowledgments}

\bibliography{references}
\appendix
\section{One-dimensional multi-scale analysis: Bulk viscosity}
Using a Taylor expansion around $(r,t)$:
\begin{equation}
	    \delta t \mathcal{D}_t f_i + \frac{{\delta t}^2}{2}{\mathcal{D}_t}^2 f_i + {O}({\delta t}^3)= 2\beta\left(f_i^{\rm eq} - f_i\right),
\end{equation}
where we have only retained terms up to order two and $\mathcal{D}_t = \partial_t + c_i\partial_x$. Then introducing characteristic flow size $\mathcal{L}$ and velocity $\mathcal{U}$ the equation is made non-dimensional as:
\begin{equation}
	   \left(\frac{\delta r}{\mathcal{L}}\right) \mathcal{D}_t' f_i + \frac{1}{2}{\left(\frac{\delta r}{\mathcal{L}}\right)}^2{\mathcal{D}_t'}^2 f_i = 2\beta\left(f_i^{\rm eq} - f_i\right),
\end{equation}
where primed variables denote non-dimensional form and
\begin{equation}
	\mathcal{D}_t' = \frac{\mathcal{U}}{c}\left(\partial_t' + c_i'\partial_x'\right),
\end{equation}
where $c=\delta r/\delta t$. Assuming acoustic, i.e. $\frac{\mathcal{U}}{c}\sim 1$ and hydrodynamic, i.e. $\frac{\delta r}{\mathcal{L}}\sim\varepsilon$, scaling and dropping the primes for the sake of readability:
\begin{equation}
	\varepsilon \mathcal{D}_t f_i + \frac{1}{2}\varepsilon^2{\mathcal{D}_t}^2 f_i + {O}(\varepsilon^3)= 2\beta\left(f_i^{\rm eq} - f_i\right).
\end{equation}
Then introducing multi-scale expansions:
\begin{subequations}
    \begin{align}
        f_i &= f_i^{(0)} + \varepsilon f_i^{(1)} + \varepsilon^2 f_i^{(2)} + O(\varepsilon^3),\\
        \partial_t &= \varepsilon \partial_t^{(1)} + \varepsilon^2 \partial_t^{(2)} + O(\varepsilon^3),\\
        \partial_x &= \varepsilon \partial_x,
    \end{align}
\end{subequations}
the following equations are recovered at scales $\varepsilon$ and $\varepsilon^2$:
\begin{subequations}
	\begin{align}
		\varepsilon &: \mathcal{D}_{t}^{(1)} f_i^{(0)} = -2\beta f_i^{(1)},\\
		\varepsilon^2 &: \partial_t^{(2)}f_i^{(0)} + \mathcal{D}_{t}^{(1)} \left(1-\beta\right)f_i^{(1)} = -2\beta f_i^{(2)},
	\end{align}
	\label{Eq:CE_Eq_orders}
\end{subequations}
with $f_i^{(0)}=f_i^{\rm eq}$. Taking the moments of the Chapman-Enskog-expanded equation at order $\varepsilon$:
\begin{subequations}
    \begin{align}
	\partial_t^{(1)}\rho + \partial_x \rho u_x &= 0,\label{eq:approach2_continuity1_app}\\
	\partial_t^{(1)}\rho u_x + \partial_x \rho u_x^2 + \partial_x \rho \varsigma^2 + \partial_x P^{*}_{xx} &= 0.\label{eq:approach2_NS1}
    \end{align}
\end{subequations}
At order $\varepsilon^2$ the continuity equation is:
\begin{equation}
	\partial_t^{(2)}\rho = 0.\label{eq:approach2_continuity2}
\end{equation} 
Summing up Eqs.~\ref{eq:approach2_continuity1_app} and \ref{eq:approach2_continuity2} we recover the continuity equation as:
\begin{equation}
	\partial_t \rho + \partial_x \rho u_x + \mathcal{O} (\varepsilon^3) = 0,
\end{equation}
For the momentum equations we have:
\begin{equation}\label{eq:eps2_mom1_1}
	\partial_t^{(2)}\rho u_x
	+ \partial_x \left(1-\beta\right)\Pi_{xx}^{(1)} = 0.
\end{equation}
where after some algebra the non-equilibrium stress tensor $\Pi_{xx}^{(1)}$ for any diagonal equilibrium pressure of the form:
\begin{equation}
    P_{\alpha\alpha} = \rho \varsigma^2 + P^{*}_{\alpha\alpha},
\end{equation}
results in:
\begin{equation}\label{eq:noneq_second_order_moment_app}
    \Pi_{xx}^{(1)} = -\frac{1}{2\beta} \left[2 A \rho \varsigma^2 \partial_x u_x + B \partial_x \rho\right],
\end{equation}
with
\begin{equation}\label{eq:general_bulk_1d_app}
    A = \left(1-\frac{3}{2}\frac{u_x^2}{\varsigma^2} - \frac{3}{2\rho \varsigma^2}u_x \partial_{u_x} P^{*}_{xx} - \frac{{(\partial_{u_x} P^{*}_{xx})}^2}{2\rho^2 \varsigma^2} - \frac{\partial_{\rho} P^{*}_{xx}}{2 \varsigma^2}\right),
\end{equation}
and
\begin{equation}\label{eq:general_compressibility_1d_app}
    B= -3u_x\partial_{\rho} P^{*}_{xx} - \frac{\varsigma^2}{\rho}\partial_{u_x}P^{*}_{xx} - \frac{\partial_{u_x}P^{*}_{xx} \partial_{\rho} P^{*}_{xx}}{\rho} - u_x^3. 
\end{equation}
\section{Multi-dimensional case\label{app:CE_multiD}}
For $D>1$, using the same procedure as the 1-D case we have the following system of equations at different orders in $\varepsilon$:
\begin{subequations}
	\begin{align}
		\varepsilon &: \mathcal{D}_{t}^{(1)} f_i^{(0)} = -2\beta f_i^{(1)},\\
		\varepsilon^2 &: \partial_t^{(2)}f_i^{(0)} + \mathcal{D}_{t}^{(1)} \left(1-\beta\right)f_i^{(1)} = -2\beta f_i^{(2)},
	\end{align}
	\label{Eq:CE_Eq_orders_multD}
\end{subequations}
where now $\mathcal{D}^{(1)}=\partial_t^{(1)} + \bm{c}_i\cdot\bm{\nabla}$, which at order $\varepsilon$ lead to:
\begin{subequations}
	\begin{align}
	\partial_t^{(1)}\rho + \bm{\nabla}\cdot\rho \bm{u} &= 0,\label{eq:approach2_continuity1_multD}\\
	\partial_t^{(1)}\rho \bm{u} + \bm{\nabla}\cdot \rho \bm{u}\otimes\bm{u} + \bm{\nabla}\cdot \rho \varsigma^2 \bm{I} + \bm{\nabla}\cdot \bm{P}^{*} &= 0.\label{eq:approach2_NS1_multD}
	\end{align}
\end{subequations}
where the tensor $\bm{P}^{*}$ is diagonal with elements $P^{*}_{\alpha\alpha}$. As for the 1-D case at order $\varepsilon^2$ the continuity equation is:
\begin{equation}
	\partial_t^{(2)}\rho = 0.\label{eq:approach2_continuity2_multD}
\end{equation} 
For the momentum equations we have:
\begin{equation}\label{eq:eps2_mom1_1_multD}
	\partial_t^{(2)}\rho \bm{u}
	+ \bm{\nabla}\cdot\left(1-\beta\right)\Pi_{2}^{(1)} = 0,
\end{equation}
with:
\begin{equation}
    \Pi_{2}^{(1)} = -\frac{1}{2\beta}\left[\partial_t^{(1)}\Pi_{2}^{(0)} + \bm{\nabla}\cdot\Pi_{3}^{(0)}\right].
\end{equation}
with:
\begin{equation}
    \Pi_{2}^{(1)} = -\frac{1}{2\beta}\left[\partial_t^{(1)}\Pi_{2}^{(0)} + \bm{\nabla}\cdot\Pi_{3}^{(0)}\right].
\end{equation}
where summation over $\gamma$ is implied. After some algebra one ends up with:
\begin{multline}\label{eq:2nd_order_moms_multiD}
    \partial_t^{(1)}\Pi_{\alpha\beta}^{(0)} = -\partial_\gamma \rho u_\alpha u_\beta u_\gamma \\ - u_\alpha \partial_\beta (P_{\beta\beta}^*+\rho \varsigma^2) - u_\beta \partial_\alpha (P_{\alpha\alpha}^*+\rho \varsigma^2) \\ + \delta_{\alpha\beta} \partial_t^{(1)}(P_{\alpha\alpha}^*+\rho \varsigma^2),
\end{multline}
while for the third-order equilibrium moments:
\begin{multline}\label{eq:3rd_order_moms_multiD}
    \partial_\gamma \Pi^{(0)}_{\alpha\beta\gamma} = \partial_\gamma \rho u_\alpha u_\beta u_\gamma + \left[\partial_\gamma u_\gamma (\rho \varsigma^2 \delta_{\alpha\beta}+P^*_{\alpha\beta})\right]_{\rm perm} \\ - \delta_{\alpha\beta\gamma} ( \left[\partial_\gamma u_\gamma P_{\alpha\beta}^*\right]_{\rm perm} 
    +  \partial_\gamma\rho u_\alpha u_\beta u_\gamma) \\ - E_{\alpha\beta\gamma},
\end{multline}
where $E_{\alpha\beta\gamma}$ is a term representing errors specific to the second order polynomial equilibrium. For both product form and entropic equilibria $E_{\alpha\beta\gamma}=0$ while for the second order polynomial equilibrium:
\begin{equation}
    E_{\alpha\beta\gamma} = (1-\delta_{\alpha\beta\gamma})\delta_{\alpha\beta}\left(\partial_\gamma \rho u_\alpha u_\beta u_\gamma + \partial_\gamma u_\gamma P_{\alpha\beta}^*\right).
\end{equation}
Combining Eqs.~\eqref{eq:2nd_order_moms_multiD} and \eqref{eq:3rd_order_moms_multiD}:
\begin{equation}
    \Pi_2^{(1)} = -\frac{1}{2\beta}\left(\bm{S}^* + \bm{D}^*\right),
\end{equation}
with:
\begin{equation}\label{eq:shear_multi_D}
    S_{\alpha\beta}^* = (\rho\varsigma^2+P^*_{\alpha\alpha})\partial_\alpha u_\beta + (\rho\varsigma^2+P^*_{\beta\beta})\partial_\beta u_\alpha  + E^s_{\alpha\beta},
\end{equation}
and:
\begin{multline}\label{eq:bulk_multi_D}
    D_{\alpha\alpha}^* = -3\left(\rho u_\alpha^2 + P_{\alpha\alpha}^* + u_\alpha \partial_{u_\alpha}P_{\alpha\alpha}^* + \frac{{\left(\partial_{u_\alpha} P_{\alpha\alpha}^*\right)}^2}{3\rho}\right)\partial_\alpha u_\alpha\\
    +\left(P_{\alpha\alpha}^* - \rho \partial_\rho P_{\alpha\alpha}^* \right)\partial_\gamma u_\gamma \\
    - \left( u_\alpha^3 + 3 u_\alpha \partial_\rho P_{\alpha\alpha}^* + \frac{\varsigma^2}{\rho} \partial_{u_\alpha}P_{\alpha\alpha}^* + \frac{1}{\rho} \partial_\rho P_{\alpha\alpha}^* \partial_{u_\alpha} P_{\alpha\alpha}^*\right) \partial_\alpha \rho\\
    + E^d_{\alpha\beta}.
\end{multline}
Here:
\begin{multline}\label{eq:2nd_order_poly_err_multiD}
    E^s_{\alpha\beta} = -\left(P^*_{\alpha\alpha}+\rho u_\alpha^2\right)\partial_\alpha u_\beta - \left(P^*_{\beta\beta}+\rho u_\beta^2\right)\partial_\beta u_\alpha \\ 
    - u_\alpha^2 u_\beta \partial_\alpha \rho - u_\beta^2 u_\alpha \partial_\beta \rho - 2 \rho u_\alpha u_\beta (\partial_\alpha u_\beta + \partial_\beta u_\alpha),
\end{multline}
and
\begin{multline}\label{eq:2nd_order_poly_comperr_multiD}
    E^d_{\alpha\alpha} = 3\left(\rho u_\alpha^2 + P_{\alpha\alpha}^* + u_\alpha \partial_{u_\alpha}P_{\alpha\alpha}^* + \frac{{\left(\partial_{u_\alpha} P_{\alpha\alpha}^*\right)}^2}{3\rho}\right)\partial_\alpha u_\alpha\\
    - \left(P_{\alpha\alpha}^* - \rho \partial_\rho P_{\alpha\alpha}^* \right)\partial_\gamma u_\gamma \\
    + \left( u_\alpha^3 + 3 u_\alpha \partial_\rho P_{\alpha\alpha}^* + \frac{\varsigma^2}{\rho} \partial_{u_\alpha}P_{\alpha\alpha}^* + \frac{1}{\rho} \partial_\rho P_{\alpha\alpha}^* \partial_{u_\alpha} P_{\alpha\alpha}^*\right) \partial_\alpha \rho\\
    + 8 \rho u_\alpha^2\partial_\alpha u_\alpha + 2 u_\alpha^3 \partial_\alpha \rho - \rho u_\alpha^2 \partial_\gamma u_\gamma - u_\alpha^2 u_\gamma \partial_\gamma \rho - 2 u_\alpha u_\gamma \partial_\gamma u_\alpha,
\end{multline}
where summation over $\gamma$ is implied in Eqs.~\eqref{eq:bulk_multi_D}, \eqref{eq:2nd_order_poly_err_multiD} and \eqref{eq:2nd_order_poly_comperr_multiD}.
Note that in 1-D, the diagonal terms in Eqs.~\eqref{eq:shear_multi_D} and \eqref{eq:bulk_multi_D} reduce to Eqs.~\eqref{eq:general_bulk_1d_app} and \eqref{eq:general_compressibility_1d_app}.
\section{Jacobian of entropic equilibrium\label{app_jacobian}}
Re-writing the entropic equilibrium as:
\begin{equation}
f_i^{\rm eq} = w_i \rho \prod_{\alpha=x,y} \Lambda_\alpha \Gamma_{i\alpha},
\end{equation}
the Jacobian can be expressed as:
\begin{equation}
\frac{\partial f_i^{\rm eq}}{\partial f_j} = f_i^{\rm eq} \left(\frac{\partial \ln \rho}{\partial f_j} + \sum_{\alpha=x,y} \frac{\partial \ln \Lambda_\alpha}{\partial f_j} + \frac{\partial \ln \Gamma_{i\alpha}}{\partial f_j}\right),
\end{equation}
with:
\begin{subequations}
	\begin{align}
	\Lambda_\alpha &= 2 - S_\alpha,\\
	\Gamma_{i\alpha} &= {\left(\frac{2u_\alpha+S_\alpha}{1-u_\alpha}\right)}^{c_{i\alpha}},\\
	S_\alpha &= \sqrt{{(u_\alpha/\varsigma)}^2+1},
	\end{align}
\end{subequations}
and :
\begin{subequations}
	\begin{align}
    \frac{\partial S_\alpha}{\partial f_j} &= \frac{1}{\rho}\left[ \frac{c_{j\alpha}u_\alpha/\varsigma^2 + 1}{S_\alpha} - S_{\alpha}\right],\\
	\frac{\partial \Lambda_\alpha}{\partial f_j} &=  -\frac{1}{\rho}\left[ \frac{c_{j\alpha}u_\alpha/\varsigma^2 + 1}{S_\alpha} - S_{\alpha}\right],\\
	\frac{\partial \Gamma_{i\alpha}}{\partial f_j} &= c_{i\alpha} \left[\frac{2c_{i\beta}+S_\alpha+\frac{\partial S_\alpha}{\partial f_j}}{\rho\left(1-u_\alpha\right)} - \frac{\left(2 u_\alpha+S_\alpha\right)\left(1-c_{i\beta}\right)}{\rho{\left(1-u_\alpha\right)}^2}\right] \nonumber\\ & {\left(\frac{u_\alpha+S_\alpha}{1-u_\alpha}\right)}^{c_{i\alpha}-1}.
	\end{align}
\end{subequations}

\end{document}